\documentclass{article}
\usepackage{authblk}

\usepackage{subfig}
\usepackage[utf8]{inputenc}
\usepackage[comma]{natbib}
\usepackage{url}
\usepackage{fullpage}
\usepackage{url}
\usepackage{xcolor}
\usepackage[colorlinks=true,
	linkcolor=violet,
	citecolor=black]{hyperref}

\usepackage{amssymb, amsfonts, amsmath}
\usepackage{graphicx, natbib}
\usepackage{algorithm2e}

\newcommand{\BIT}{\begin{itemize}}
\newcommand{\EIT}{\end{itemize}}
\newcommand{\BNUM}{\begin{enumerate}}
\newcommand{\ENUM}{\end{enumerate}}

\newcommand\mbb[1]{\mathbb{#1}}

\def\mrm#1{\mathrm{#1}}
\def\reals{\mathbb{R}} 
\renewcommand{\exp}[1]{\operatorname{exp}\left(#1\right)} 
\def\indic#1{\mbb{I}\left({#1}\right)} 

\def\E{\mathbb{E}} 
\def\Earg#1{\E\left[{#1}\right]}
\def\Esubarg#1#2{\E_{#1}\left[{#2}\right]}
\def\P{\mathbb{P}} 

\def\Var{\mrm{Var}}
\def\Vararg#1{\Var\left(#1\right)}

\def\absarg#1{\left|#1\right|}

\def\Gsn{\mathcal{N}}
\def\Ber{\textnormal{Ber}}

\def\Mult{\textnormal{Mult}}

\newcommand{\Scal}{\mathcal{S}}

\def\*#1{\mathbf{#1}}

\DeclareGraphicsExtensions{.pdf, .jpg, .gif,.png,.bmp}

\title{Generative Models: An Interdisciplinary Perspective}

\author[1]{Kris Sankaran}
\author[2]{Susan P Holmes}
\affil[1]{Statistics Department, University of Wisconsin - Madison, Madison, USA, WI 53706; ksankaran@wisc.edu}
\affil[2]{Statistics Department, Stanford University, Stanford, USA, CA 94305; susan@stat.stanford.edu}

\begin{document}
\maketitle
\begin{abstract}
By linking conceptual theories with observed data, generative models can support reasoning in complex situations. They have come to play a central role both within and beyond statistics, providing the basis for power analysis in molecular biology, theory building in particle physics, and resource allocation in epidemiology, for example. We introduce the probabilistic and computational concepts underlying modern generative models and then analyze how they can be used to inform experimental design, iterative model refinement, goodness-of-fit evaluation, and agent-based simulation. We emphasize a modular view of generative mechanisms and discuss how they can be flexibly recombined in new problem contexts. We provide practical illustrations throughout, and code for reproducing all examples is available at \href{https://github.com/krisrs1128/generative_review}{https://github.com/krisrs1128/generative\_review}. Finally, we observe how research in generative models is currently split across several islands of activity, and we highlight opportunities lying at disciplinary intersections.
\end{abstract}
\begin{keywords}
generative models, simulation, decision-making, agent-based models, experimental design, particle filter, model evaluation, goodness-of-fit
\end{keywords}

\maketitle

\section{Introduction}


In the classroom, a simple trick for teaching complex statistical ideas is to first generate a dataset from a familiar model. A simulation can turn a problem of logic into one of observation. For example, to introduce the concept of a Gaussian mixture model, one approach is to write a marginal density,
\begin{align*}
p\left(x\right) = \pi\varphi\left(\frac{x - \mu_1}{\sigma_1}\right) + \left(1 - \pi\right)\varphi\left(\frac{x - \mu_2}{\sigma_2}\right),
\end{align*}
but, following \citep{HolmesHuber}, it is much more evocative to give a demonstration,
\begin{enumerate}
\item Flip a coin, $z_{i} \sim\Ber\left(\pi\right)$
\item Simulate,
	 \begin{align*}
		x_{i} \vert z_{i} \sim \begin{cases}
		\Gsn\left(\mu_1, \sigma_1^2\right) &\text{ if }z_{i} \text{ comes up heads}\\
		\Gsn\left(\mu_2, \sigma_2^2\right) & \text{ otherwise}.
		\end{cases}
		\end{align*}
\item Repeat this for $i = 1, \dots, N$ and make a histogram of $x_i$.
\end{enumerate}
This is a simple setting, but it already exhibits some salient features of generative models. First, a more complex mechanism is composed from simple building blocks. Second, it invites experimentation, and we can easily generate datasets with, say, different biases $\pi$. Further, it is “top down” in the sense that we specify latent variables and parameters before observing $x_{i}$. Finally, it makes model evaluation natural, since both the model and real data can be understood in terms of histograms in the data space.

This capacity to clarify has been recognized since even before the widespread use of computers — see for example the generation of mixtures in \citep{Teicher1960mixture} or the language of particles in \citep{metropolis1949monte} at the onset of the Monte Carlo revolution. These early efforts have matured into methods that enable simulation and inference in very general settings. For example, MCMC and Gibbs sampling have made it possible to work with models whose normalizing constants are unknown \citep{diaconis2009markov}, approximate Bayesian modeling supports inference even when likelihoods are inaccessible \citep{beaumont2019approximate}, and measures of distributional discrepancy enable iterative model refinement \citep{anastasiou2021stein}. Generative models now play a central role throughout design of experiments, data modeling, inference (with or without likelihoods), and decision-making under uncertainty.


Generative models are also invaluable communication tools. For instance, it can be difficult to raise identifiability concerns with collaborators, especially when the statistical analyses require specific technical knowledge. However, a simulation showing how alternative scenarios can result in indistinguishable outputs makes non-identifiability clear \citep{brun2001practical}. Indeed, this idea has taken on a life of its own, resulting in a literature outside statistics on “sloppy” (i.e., nonidentifiable) models \citep{gutenkunst2007universally}.

Modern generative models relieve the researcher from needing full distributional specifications. There are now procedures combining optimization with simulation to minimize discrepancies between model and data, even when complete densities are unavailable. For example, \cite{gutmann2016bayesian} develop an adaptive modeling strategy based on Bayesian optimization. More generally, the gap between statistical and simulation modeling is narrowing, with domain-specific simulators in single-cell genomics and high-energy physics beginning to make contact with ideas like optimal experimental design \citep{sun2021scdesign2} and the Neyman-Pearson lemma \citep{dalmasso_likelihood-free_2021}, for instance.

At the other end of the specification spectrum, many scientific communities, ranging from economics and sociology to biology and physics, have refined bottom-up generative models based on theoretically-guided rules. These models, often called Agent Based Models (ABMs), simulate complex systems by providing rules that govern interactions across large numbers of similar agents \citep{an2009agent, grazzini2017bayesian}. Such models have been used for modeling disease dynamics, and we will show how they can be used to infer mutation rates in butterflies and have assisted policy makers during the COVID-19 pandemic.

Our review follows the timeline of a statistical workflow: starting in section \ref{sec:design} we use generative models for power computations and experimental design, guiding decision-makers through a choice about blocking designs, and outlining the dense landscape of R packages that use generative models for power and design computations. Section \ref{sec:models} covers model building from the perspective of a grammar of generative models, expressing complex dependencies by composing elementary stochastic modules. Section \ref{sec:calibration} reviews goodness-of-fit and discrepancy measures, and we provide an example of using statistical learning methods to partially automate the process of evaluating a model’s approximation quality. In section \ref{sec:emulation}, we show several examples where generative models enable the study of complex systems, like adaptation of a population through natural selection and the control of an epidemic outbreak. Finally, the conclusion outlines ideas that are driving current progress in the field, which have their origins across many communities. We highlight the interdisciplinary, future challenges in generative modeling that will benefit scientists and policymakers.



\section{Experimental Design}
\label{sec:design}

A wise choice of experimental design can deliver powerful inferences even when resources are limited. Conversely, a careless approach can make some inferences impossible, no matter the analysis strategy. For this reason, the design of experiments has a long history in statistics \citep{fisher1937design, cox2000theory}. However, the emergence of studies where data are collected from a variety of populations, environments, or sensing devices, has prompted further development. Indeed, the increasing ease in scientific communication and data sharing have supported the implementation of studies of unprecedented scope. In this section, we review some challenges of planning modern studies and the ways generative modeling has become integrated into experimental design.

Regardless of the application context, a few pressing questions arise in any experimental plan,
\begin{itemize}
    \item How effectively will we be able to measure a quantity of interest? Are there alternative designs that would improve power?
	\item Are there certain competing scientific scenarios that will be impossible to disambiguate given the data generated?
	\item How many samples will be enough to achieve a certain degree of confidence in the strength and directions of effects?
	\item Given a fixed budget limit, how should controllable factors be varied?
\end{itemize}
To answer these questions systematically, classical experimental design arrived at a few core principles, like randomization, blocking, factorial, and sequential design. These concepts provide a language for expressing important considerations in any design. However, they do not necessarily guide the planner towards any particular choice. Additional criteria are needed for (i) evaluating the utility of a design and (ii) navigating the space of possibilities. Classical experimental design provides the former, generative models the latter. Notions of optimality are defined with respect to a model fitting procedure. For example, suppose a model is fit using $N$ design points $x_{1:N}$, and denote the variance of the associated prediction surface at configuration $x$ by $\text{Var}\left[\hat{y}_{x_{1:N}}\left(x\right)\right]$.
Then, a design $x^\ast_{1:N}$  minimizing $\int_{\mathcal{E}} \text{Var}\left[\hat{y}_{x_{1:N}^\ast}\left(x\right)\right]dx$ over the set of all candidate designs $x \in \mathcal{E}$ is called $I$-optimal. A number of other optimality criteria have been proposed, see for example \citep{john1975d, cox2000theory}.

These quantitative criteria measure efficiency changes induced by particular design choices, e.g., blocking or sample size.
For simple  models and experimental universes $\mathcal{E}$, it is possible to derive optimal designs in closed form. For example, when $\mathcal{E} = \left[-1, 1\right]^{K}$ and linear regression is used, then a budget of $N = 2^{K}$ samples is most efficiently used by placing all points at corners of the cube $\left[-1, 1\right]^{K}$. In more complex settings, however, analogous mathematical results may not be available.

\subsection{Challenges in modern experimental design}

To guide selection of a design, we can simulate from plausible generative models and empirically estimate the efficiency of competing designs. Before discussing this approach in detail, we note the sources of complexity in modern design.

Often, modern studies focus on systems with many interacting components, and investigators hope to attribute observed properties to subtle structure within these systems. For example, in biology, it has become increasingly common to conduct multi-domain studies, gathering multiple, complementary assays for each sample under consideration \citep{bolyen2019reproducible, lahnemann2020eleven}. Counts of microbial species, called taxa, may be studied in conjunction with metabolomic and host gene expression profiles and even stained in-situ images. In this setting, different experimental treatments may affect both individual system components and their interactions, any of which may have consequences for host health. Adding to the difficulty of characterizing these multi-component systems is the fact that studies increasingly supply data across multiple scales and modalities. For example, in ecology, a climate model may provide biome-level forecasts, a satellite product may highlight kilometer-level summaries, and ground sensor measures may provide precise, but localized data. Using these data to support climate change adaptation requires integration across all sources \citep{diffenbaugh2020covid, schmitt2016data}.

This high degree of data heterogeneity creates difficulties for characterizing optimal designs. Compounding this challenge is the diversity of modern sampling and analysis strategies. For example, in longitudinal studies, experimenters must choose sampling times, and in multidomain studies, the choice of assays per subject can be adjusted. In some cases, multiple treatments and populations may be studied simultaneously. Indeed, interest may lie in differential treatment effects across sub-populations. Moreover, there may be a trade-off between high-quality, but expensive samples, and low-quality, but abundant proxies.

\subsection{Simulation modeling for experimental design}

At first, these complexities might seem insurmountable. However, it is often possible to guide decisions using simulation, and realistic generative mechanisms can provide precise material for decision-making. To compare sampling strategies, it is valuable to imagine hypothetical experimental outcomes and their subsequent data analysis results \citep{muller2005simulation, huan2013simulation, zhang2019frequentist}. This provides an iterative, adaptive alternative to the fixed, universal strategies common in classical experimental design,
\begin{enumerate}
\item Propose a generative mechanism $p\left(x; e, \eta\right)$ that simulates a range of plausible, hypothetical datasets $x$ given a design $e \in \mathcal{E}$ and generative parameters $\eta \in H$.
\item Implement an initial version of the analysis. Quantify the utility of the analysis with a function $U\left(x, e, \eta\right)$, and investigate how design and simulation choices influence the expected utility.
\item Explore the design space $e \in \mathcal{E}$ to identify regions of high expected utility, and refine the generative mechanism to provide structures $x$ similar to those seen in earlier studies.
\end{enumerate}
We first illustrate this process using two concrete design problems. We then survey how more advanced versions of this approach have proliferated through application-specific packages and workflows.

\subsubsection{Blocking in generative count models}

Imagine we are designing a study whose purpose is to evaluate the influence of three interventions on microbiota community composition. For example, one  intervention might be a diet change, and we hope to identify taxa whose abundances are systematically elevated or depressed following the intervention. The task is to guide the choice of the number of samples and the allocation of treatments across study participants. Previous studies point to substantial subject-to-subject variation in taxa composition. For this reason, a blocked design seems appropriate, where each subject is assigned each of the three interventions in random order. We argue that, by allowing estimation of a subject-level baseline, such a design will allow detection of small treatment effects even in the presence of large subject-to-subject differences. To our surprise, we face resistance to this proposal — we are told that asking each study participant to contribute three samples creates an undue burden and risks increased study dropout.

To sharpen an analysis of the trade-offs involved, a simulation can be used. Suppose that the biological specimen,  sample $i$’s community composition, is drawn from
\begin{align}
\label{eq:blocking_model}
\begin{split}
x_{i} \vert p_{i} &\sim \Mult\left(N_{i}, p_{i}\right) \\
p_{i} &\sim \Scal\left(\beta_{t_{i}} + \tau_{s_{i}}\right) \\
\beta_{t, A} &\sim \Gsn\left(0, \sigma^2_{\beta}\begin{pmatrix} I_{\absarg{A}} & 0 \\ 0 & 0_{\absarg{A^C}} \end{pmatrix}\right) \\
\tau_{s} &\sim \Gsn\left(0, \sigma^2_{\tau}I_{D}\right)
\end{split}
\end{align}
where $\Scal$ is the softmax function, $\Scal\left(z\right) = \frac{\exp{z_{d}}}{\sum_{d^\prime = 1}^{D} \exp{z_{d^\prime}}}$, and $\beta_{t}$ and $\tau_{s}$ encode treatment ($t = 1, 2, 3$) and subject effects across the $D$ species, respectively. Let $A$ designate the set of taxa that exhibit any treatment effect. Whenever $\sigma^2_{\tau} \gg \sigma^2_{\beta}$, it is clear that blocking is necessary; conversely, when $\sigma^2_{\tau} \ll \sigma^2_{\beta}$, there is no reason to create an extra burden on study participants. But what is the appropriate decision in more intermediate cases?

We study the question by simulating from the model (\ref{eq:blocking_model}) under both blocked and unblocked  designs and then sample from the posteriors $p\left(\beta \vert x_{1:N}\right)$ using Markov Chain Monte Carlo (MCMC). We set $N = 60$ as the total number of samples (60 subjects in the unblocked design, 20 for the blocked design) and $D = 25$  features. We suppose that that $\sigma_{\beta}^2 = 0.5$ and $\sigma^2_{\tau} = 2$ and that only features $d = 1, \dots, 10$ contain true effects. 

\begin{figure}
    \centering
    \includegraphics[width=0.9\textwidth]{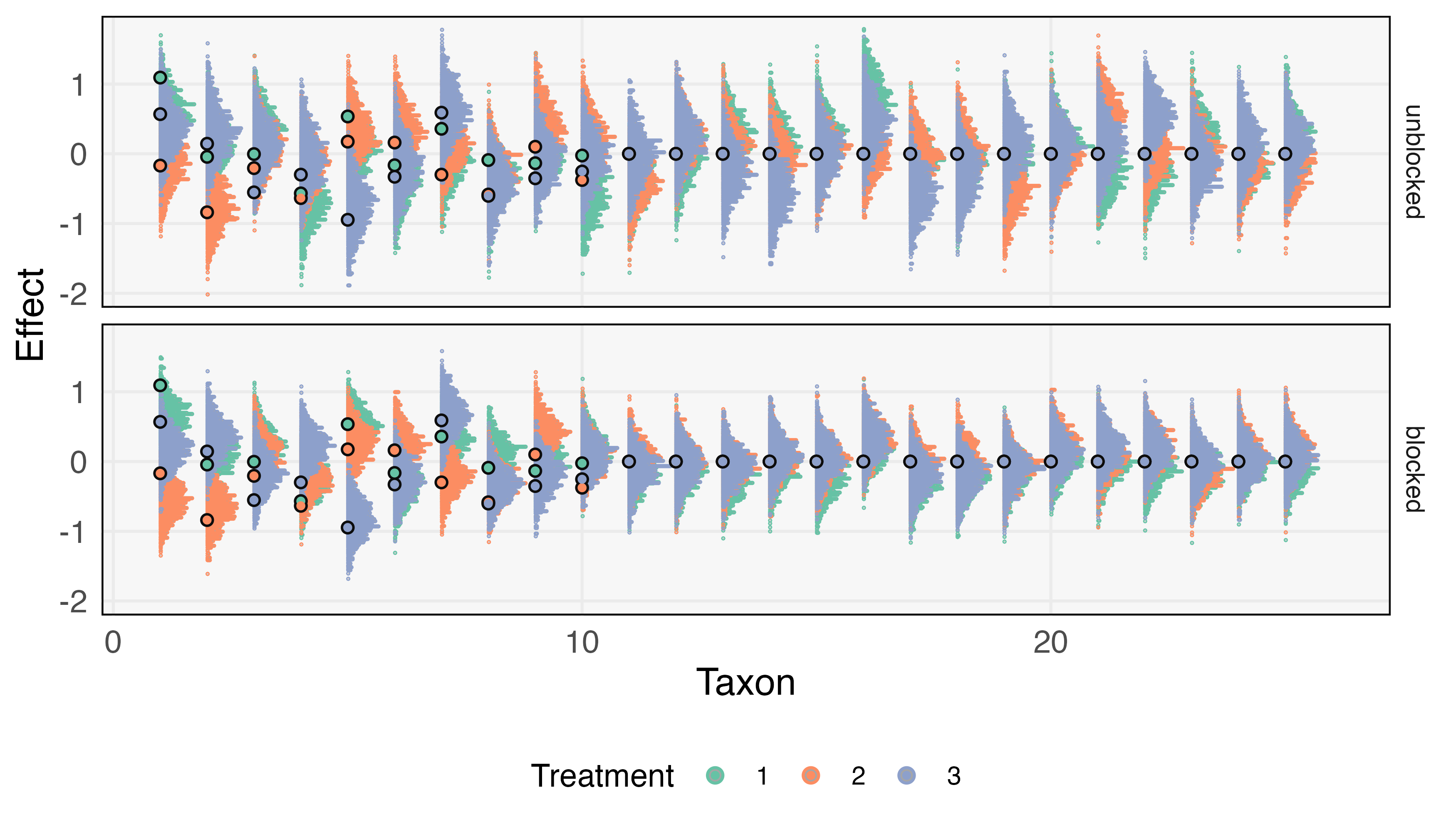}
    \caption{A comparison of posterior treatment effects estimated using a unblocked (top row) and blocked (bottom row) designs. Each column corresponds to one taxon. Circles mark the true treatment effects. Note that only the first 10 of 25 taxa have any nonzero treatment effects.}
    \label{fig:blocking_counts}
\end{figure}

Figure \ref{fig:blocking_counts} shows the resulting posteriors. Both designs support recovery of the direction for each treatment effect and are shrunken towards 0, a natural consequence of the model’s hierarchical structure. However, for each taxon $d$, we note that the blocked design has narrower posteriors $p\left(\beta_{td} \vert x_{1:N}\right)$. Specifically, the marginal variances under the unblocked design are on average 1.3 times larger than those under the blocked design. Practically, this leads to reduced power against weak effects in the unblocked design, consider for example the difference in estimated effects for taxon 8. Perhaps even more troubling are the instances in the unblocked design where we might be misled into believing in the existence of a treatment effect; consider for instance, taxon 19 or 21. This is likely the consequence of a few subjects having large (or small) abundances for these taxa simply by chance, and then dragging their corresponding treatments up (or down). Though the unblocked design gives theoretically unbiased inferences, in finite samples we still risk spurious associations. The repeated subject-level samples in the blocked design allow these outlier taxa-subject combinations to be directly accounted for.

Overall, blocking seems natural given this hypothetical setup. However, our point  is less the final choice of design than the comparison process. If experience had suggested a different ratio of $\sigma^2_{\beta}$ to $\sigma^2_{\tau}$, the same process may have led us to the opposite decision. Moreover, this simple analysis could be enriched in several directions. It is possible to model differential study dropout in the two designs. If data were available from studies with similar expected treatment effects, we could tailor the sizes of $\sigma^2_{\beta}$ and $\sigma^2_{\tau}$ to more closely match past data. Finally, an interactive version of Figure \ref{fig:blocking_counts} would support rapid comparison across simulation parameters.

\subsubsection{Sample allocation in longitudinal studies}

Even though the generative mechanism in the previous example is complex, the associated design space was simple — the choice was between either a blocked or an unblocked design. In more complex design spaces, it is impractical to evaluate all candidate designs. In this setting, an organized search over candidates can allow the more efficient discovery of promising designs \citep{muller2005simulation, huan2013simulation}. To make this search possible, we assume that utility tends to vary smoothly over the design space. Designs similar to those that are known to be ineffective can be rapidly discarded, and exploration can be concentrated on areas of higher expected utility or uncertainty.

\begin{figure}
    \centering
    \includegraphics[width=\textwidth]{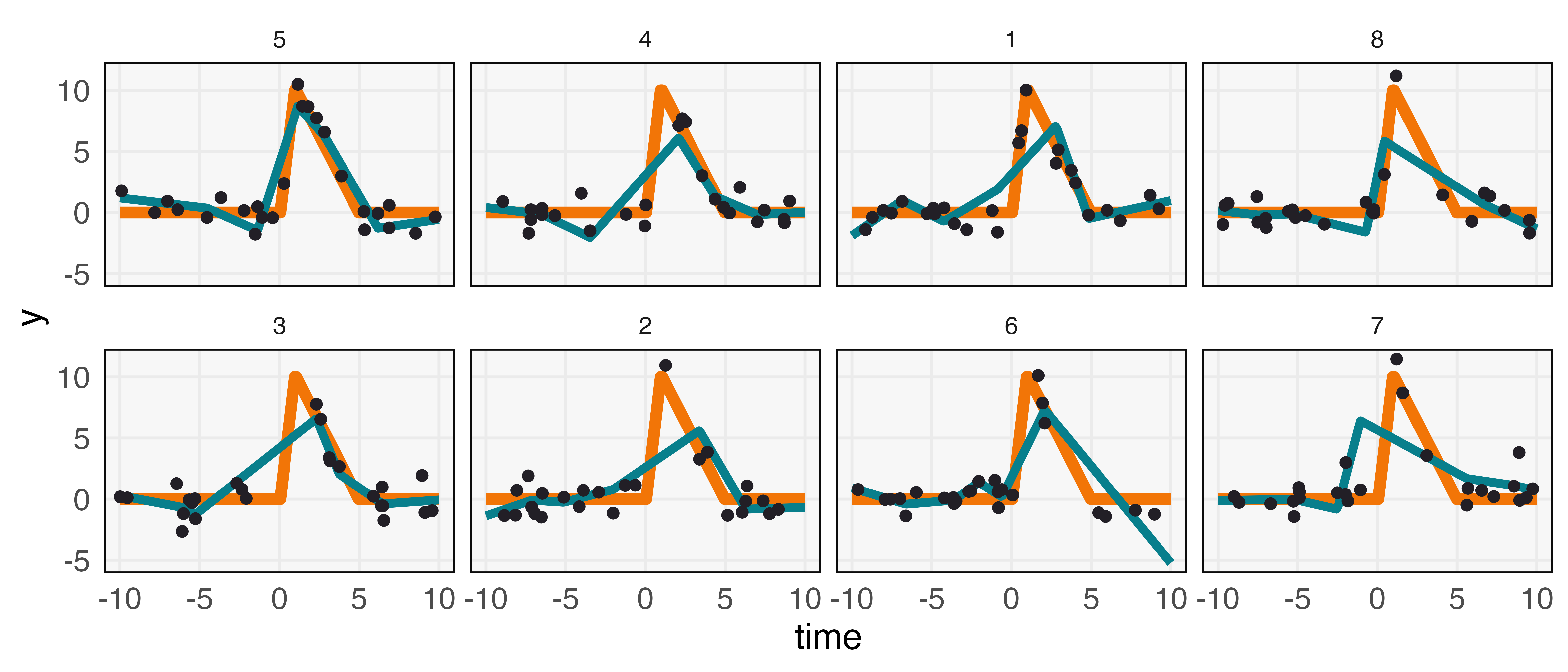}
    \caption{Example longitudinal fits from data simulated according to random designs. The solid orange curve is the unknown intervention effect, and the green curves show spline fits based on the data from that panel’s run. Panels are sorted from those with the lowest to those with the highest integrated mean squared error.}
    \label{fig:peak_samples}
\end{figure}

To illustrate, consider the problem of designing an experiment to evaluate the short and long-term effects of an intervention; e.g., the effect of antibiotic treatment on a bacterial strain of interest. The intervention effect is thought to be strong but brief, see Figure \ref{fig:peak_samples}. Sampling is costly, and in the absence of a formula for the optimal sample placement, we turn our attention to simulation.

In principle, the design space can consist of the locations of the $N = 25$ sampling points directly. However, searching across $\left[-10, 10\right]^{N}$ becomes difficult for even moderate $N$. Instead, we suppose that sampled timepoints are drawn from $\P_{w} := \sum_{k = 1}^{K} w_{k} \indic{S_{k}}$, a mixture of uniforms over predefined sets $S_{k}$ (here, we partition $\left[0, 1\right]$ into $K = 7$ equal length intervals) but with weights to be optimized. In this way, we reduce the search to the $K - 1$-dimensional simplex.

Suppose that $f$ is the true effect and $\hat{f}_{x}$ is our estimate based on a specific sampling times $x$. We write the  expected loss for a given set of weights by $L\left(w\right) := \Esubarg{x\sim \P_{w}}{\int_{\left[-10, 10\right]}\left(\hat{f}_{x}\left(z\right) - f\left(z\right)\right)^2 dz}$. This is an ideal loss function that can be used to search over the simplex; it can be empirically approximated by sampling $B$ different designs $x^{b}$ from a proposed $\P_{w}$, computing an estimate $\hat{f}_{x^b}$ (we use a spline with six degrees of freedom), and then computing $\frac{1}{n_{z}}\sum_{b = 1}^{B} \sum_{i = 1}^{n_z} \left(\hat{f}_{x^b}\left(z_{i}\right) - f_{x^b}\left(z_i\right)\right)^2$ across a dense grid $z_i \in \left[-10, 10\right]$.

The loss $L\left(w\right)$ can be used to compare alternative designs $w$, but evaluating it across all of $\Delta^{K - 1}$ is computationally impractical. As an alternative, we apply Bayesian optimization to iteratively guide our search towards promising $w$ \citep{shahriari2015taking}. This approach first places a Gaussian Process prior on the loss function $L\left(w\right)$. This encodes the belief that no particular configuration of $w$ should be favored a priori, but that the loss should vary smoothly as a function of $w$. The optimization strategy balances exploration and exploitation of competing designs, narrowing in on promising configurations without prematurely ruling out alternatives. Specifically, we first randomly sample an initial set of weights $w_{1}, \dots, w_{N_{\text{init}}}$ from a logistic-normal distribution. For each weight $w_{i}$, we evaluate $L\left(w_i\right)$. The paired $w_{i}, L\left(w_{i}\right)$ are used to compute a posterior. Given this posterior, we find $w$ with small values of $\Earg{L\left(w\right)} - \Vararg{L\left(w\right)}$, representing points with either low expected loss or high uncertainty. We evaluate $L\left(w\right)$ at these candidate $w$, update the posterior, and continue until convergence.

\begin{figure}
    \centering
    \includegraphics[width=\textwidth]{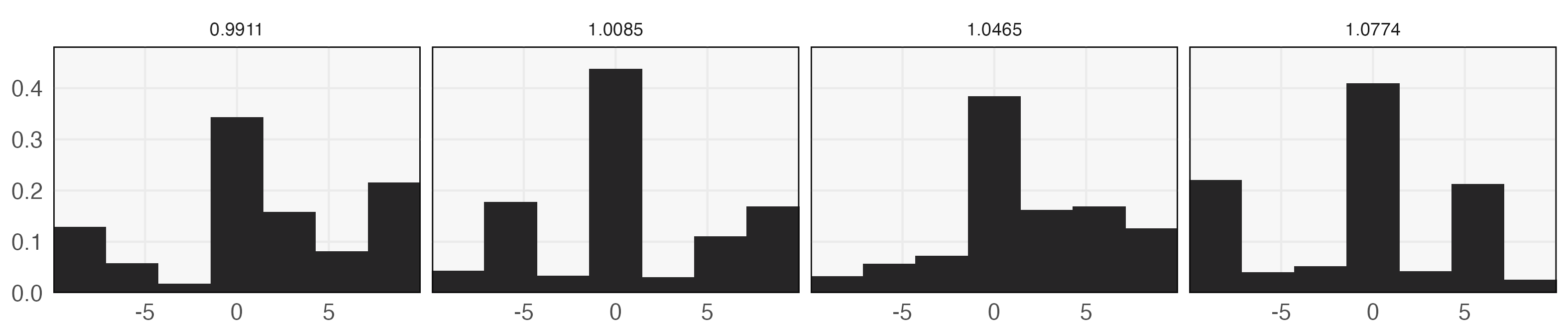}
    \caption{The four distributions $\P_{w}$ giving the lowest estimated loss $L\left(w\right)$, identified using Bayesian Optimization. Panel names give the estimated loss of the associated configuration. Comparing with Figure \ref{fig:peak_samples}, we observe that configurations place high weight in the region of the peak. All but the third configuration also place high weight near start and end times.}
    \label{fig:best_weights}
\end{figure}

A few choices of $\P_{w}$ discovered through this process are given in Figure \ref{fig:best_weights}. These distributions place elevated weight near the intervention, matching the intuition that sampling should be densest in regions where the effect is expected to vary the most. To summarize, the overall approach is,
\begin{itemize}
\item Specify a generative mechanism. In this example, it is the shape of the unknown treatment effect and the distribution of random variation around the true effect.
\item Define a design space and loss. This is necessary to formally evaluate competing designs. Here, the design space is the choice of weights in $\P_{w}$ and the loss is the average estimation error when using sampling times drawn from $\P_{w}$.
\item Optimize over designs. Iteratively evaluate losses on candidate designs, choosing candidates in a way that balances exploration with exploitation. In this example, we used Bayesian optimization.
\end{itemize}

Of the three steps, the final optimization is the most automatic. Specification of the generative mechanism and the loss require care. That said, the full modeling toolkit can be applied to improve the faithfulness of the specification, giving the designer considerable flexibility.

\subsubsection{Software ecosystems for design-by-simulation}

The examples above show how simulators can be built from scratch to support experimental design. However, for many interesting data types, there are already a number of available simulators. A complete inventory of available packages for simulation-based design of experiments is not possible, a list of R packages for Experimental Design can be found at {\tt https://cran.r-project.org/web/views/ExperimentalDesign.html}. There are in fact hundreds of packages for doing design by simulation -- we have curated a list in Supplementary Table 1. These packages make it possible to reuse code across the community, minimizing duplicated effort planning designs.

The simulators used within a community must be tailored to properties of the data it must generate. Nonetheless, for design-by-simulation, a few properties are generally useful,
\begin{enumerate}
\item Calibration: We should be able to calibrate simulation outputs to match existing datasets. Rather than simulating data de novo, public datasets and pilot studies can be used to set expectations for a new experiment.
\item Evaluation: Metrics should be available to evaluate the discrepancy between (a) real and simulated datasets and (b) analysis results derived from real and simulated data.
\item Control: We should be able to directly manipulate both experimental and generative parameters. A simulator that produces realistic data is not useful for power calculations if it can only be run for a single sample size, for example.
\item Transparency: The mechanisms leading to simulated data should be interpretable and refer to language already used within the application domain.
\item Usability: A well-documented package and interactive interface can make a simulator more accessible, supporting its adoption.
\end{enumerate}

To see how these principles can guide practice, consider how simulation packages guide experimental design in single-cell genomics. A variety of simulation mechanisms have been proposed to capture several properties common across single-cell datasets: overdispersed counts, a high degree of sparsity, low effective dimensionality, and the presence of batch effects \citep{zappia2017splatter, risso2017zinb, zhang2019simulating, yu2020scssim, sun2021scdesign2, schmid2021design, qin2022scrip}. 
Most of these simulators are built from rich probabilistic or differential equations models. For example, scDesign2 models each gene marginally using a zero-inflated negative binomial distribution and then induces correlation using a copula \citep{li2019statistical, sun2021scdesign2}. SymSin uses a sequential strategy to model the transformation of transcripts within a cell to observed counts in a single-cell dataset \citep{zhang2019simulating}. Some simulators are designed with calibration and control built-in — for example, given a pilot dataset, they make it possible to simulate a version of the data with twice as many cells but with fewer differentially expressed genes.

The community has additionally proposed a variety of metrics for evaluating the faithfulness of these simulators. For example, gene-level distributions, cross-gene mean-variance relationships, dataset-level sparsity, and comparability of derived dimensionality reduction visualizations have all been used to measure discrepancy between real and simulated data \citep{cao2021benchmark, soneson2018towards}. Both simulators and benchmarking (comparing performances in standard situations) have been encapsulated into packages for wider dissemination.

Note that these simulators were not originally designed with experimental design in mind -- instead, most were written to support benchmarking. This was necessary because manual annotation of real single-cell data requires expertise and is labor-intensive. In contrast, simulators come with “ground truth” annotations. Nonetheless, after these simulators were implemented, they began to be used for design studies, and a later generation of simulators were written explicitly with design studies in mind.

\subsection{Experimental design for simulation modeling}

Evidently, generative models can inform modern experimental design. This is only one side of the story, however — the principles of design can support the implementation of ever-more-sophisticated simulation experiments. This has been the subject of two reviews in Statistical Science, first in \citep{sacks1989design} and more recently in \citep{baker2022analyzing}.

\citet{sacks1989design} were pioneers who realized that even simulations of  deterministic physical systems,
such as fluid dynamics and thermal energy storage, could benefit from stochastic models at the computational design step. In this setting, detailed knowledge of the data generating mechanism means that confounding is a less central concern, and experimental design concepts like blocking and randomization are rarely needed. However, they underline that the clever choice of inputs can lead to more efficient use of computational resources, and that these choices can be guided by experimental design principles. For example, it is beneficial to allocate more samples in regions with high variability, analogous to sequential design strategies used in response surface design and  kriging. \citet{baker2022analyzing} explore subsequent developments, demonstrating how the decision of where to evaluate stochastic simulators can be guided by the theory of space-filling and sequential designs. They emphasize recent progress in emulating the full distribution of simulation outputs, rather than just their means. They illustrate these techniques using an application to computational ocean science, where differential equation-based simulation models are widely used. 

\section{Model building}
\label{sec:models} 

It is helpful to view generative model building as a language, not simply selection from a fixed catalog. By learning the appropriate statistical vocabulary and grammatical rules, it becomes possible to express complex dependence structures between observed and latent data in a way that guides description and decision-making.
Specifically, in the way that a language is composed of words, generative models are composed of elementary stochastic modules. By linking these elements through conditional dependence relationships, it is possible to induce dependence across both samples and measurements. 

Historically, generative modeling has been criticized as unnecessarily constraining, relative to the flexibility of algorithmic modeling \citep{breiman2001statistical}. However, an effective model provides a succinct description of the processes behind an observed dataset, and the resulting summaries are much richer than predictions alone. Moreover, concerns about flexibility have been addressed by advances in the number and variety of base modules amenable to computation and inference \citep{ghahramani2015probabilistic, van2018introduction}. Longitudinal, spatial, count, zero-inflated, and multi-domain structure can now be routinely generatively modeled using open-source, accessible software libraries \citep{carpenter2017stan, bingham2019pyro, wood2014new}.

\subsection{Grammar rules}

\begin{figure}
    \centering
    \includegraphics[width=\textwidth]{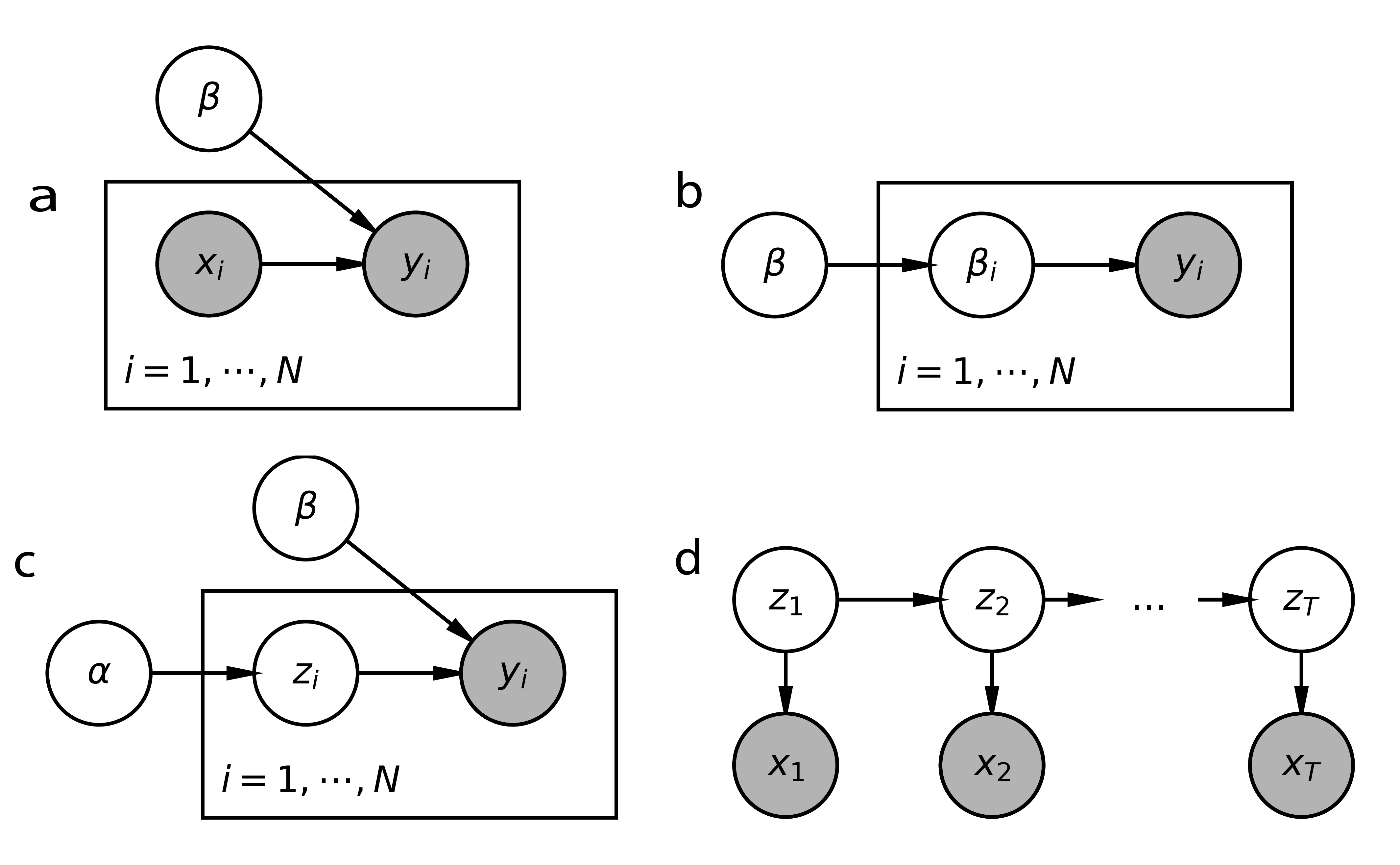}
    \caption{DAG representations of common stochastic building blocks, linear regression (a), hierarchical (b), and local latent (c) structure. Circles are shaded if the associated random variables are observed. A directed edge is drawn if the target variable is conditional dependent on the source. The rectangular ``plate'' in the diagram is used to represent $N$ copies of the contained variables.}
    \label{fig:lr_dag}
\end{figure}

Any generative model can be written as a sequence of (potentially conditional) sampling steps. These sampling steps can be represented mathematically, computationally, and graphically. For example, the familiar linear regression model can be equivalently written mathematically as,
\begin{align*}
y_{i} \vert \beta &\sim \Gsn\left(x_{i}^{T} \beta, \sigma^2_{y}\right) \text{ for } i = 1, \dots, N\\
\beta &\sim \Gsn\left(0, \sigma^2_{\beta}I_{D}\right)
\end{align*}
computationally as,
\begin{verbatim}
beta = random_normal(D, 0, sigma2_beta)
for (i = 1...n)
    y[i] = random_normal(N, x[i] * beta, sigma2_y)
\end{verbatim}
and graphically as Figure \ref{fig:lr_dag}a.

The set of edges in the DAG notation is determined by the conditional dependence structure in the original mathematical and computational formulations. Further, the \texttt{for} loop and the ``plate'' are notationally equivalent. Both are used to encode conditionally independent sampling, compressing what would be otherwise nearly identical lines of code or large fans in the graph, respectively. Note that DAG representation omits the specific distributions used, preserving only the set of conditional dependencies. This loses information that must be added back before formal implementation. However, it provides a convenient shorthand for developing more complex models. The compactness of the notation facilitates comparison of related models, a quick glance can reveal nodes or edges that are included in one, but not the other.

\subsection{A tour of stochastic modules}

\subsubsection{Hierarchical modeling}
When a dataset has many related, but not identically distributed, subsets, then a hierarchical structure may be applicable \citep{draper1993combining, gelman2006multilevel, hjort_hierarchical_2010}. In this type of module, a collection of parameters, one for each subset, is drawn from a shared prior. This implements a form of partial pooling — information is shared across subsets, but heterogeneous effects are reflected in the unique parameter estimates across subsets.  Moreover, this facilitates adaptive estimation, strongly shrinking noisily estimated parameters towards a global estimate while minimally affecting parameters in subsets with clear signals. The DAG representation is provide in Figure \ref{fig:lr_dag}b. 

This structure is frequently encountered in practice. For example, when estimating differential expression across a collection of genes, the true effect for each gene can be drawn from a shared prior, which may itself be adapted according to the distribution of observed test statistics \citep{stephens2017false, wang2021empirical}. When modeling multiple gene expression profiles simultaneously, a hierarchical model improves power by borrowing strength rather than fitting each profile separately \citep{stephens2013unified, flutre2013, zhao_bayesian_2016}. Similarly, in political science, county-level effects can be partially pooled, stabilizing inference in regions with few samples \citep{hill2017changing}. In more complex settings, hierarchical components may need to be embedded within larger generative models. For example, if the same counties were measured over time, we may believe each county’s parameter has the potential to change slightly. Viewed more generally, the reason this module is valuable is because large datasets rarely arise as a large collection of truly independent and identically distributed samples \citep{jordan2011open}. Instead, they tend to be an amalgamation of related, but not identical subsets — big data often arise by gluing many small datasets.

\subsubsection{Latent variables}

Modules based on latent variables provide a way to induce dependence structure across samples or measurements. They are applicable when it is thought that data lie within clusters or low-dimensional gradients, but the specific form of these structures is not known in advance. In the grammar of generative models, each sample is associated with a “local” latent variable and observations are drawn independently, conditionally on these latent variables \citep{blei2014build, zhang2009latent}. For example, in a generative clustering model, each sample is associated with a latent cluster indicator. For probabilistic factor models, the analogous local latent variables are low-dimensional coordinates, and in a topic model, they are mixed-memberships across a collection of topics \citep{nguyen2017bayesian, sankaran2018}. In contrast, the parameters defining the locations of clusters, shape of the gradient, or factor loadings are summarized by “global” variables.

Although different types of local and global variables induce very different types of observed data, all models with this type of local-global breakdown have the same DAG conditional independence structure, shown in Figure \ref{fig:lr_dag}c. As in hierarchical models, this DAG module can be embedded within larger models to account for other problem-specific structure. Of particular interest is latent structure that evolves over time, which we review next.

\subsubsection{Latent spatial and temporal structure}
\label{subsubsec:structured_latents}

In data with a spatial or temporal component, we often expect dependence between close by samples. For example, in a model of the audio recording of a meeting, a latent variable $z_{t}$ may learn to distinguish different speakers. Since each speaker will typically speak across an uninterrupted segment of time, we should encourage $z_{t}$ to only switch occasionally. One approach is through a latent Markov model, whose DAG structure is given in Figure \ref{fig:lr_dag}d. In this model, the latent variables follow Markov dynamics $p\left(z_{t + 1} \vert z_{1:t}\right) = p\left(z_{t + 1} \vert z_{t}\right)$. Given these latent states, the observed data are conditionally independent, specified by $p\left(x_{t} \vert z_{t}\right)$. 
For simple likelihoods $p\left(x_{t} \vert z_{t}\right)$ and dynamics $p\left(z_{t + 1} \vert z_{t}\right)$, this quantity can be computed through a closed-form recursion, like the Viterbi algorithm or Kalman Filter. More complex settings are discussed in Section \ref{subsubsec:pf}.
For example, assuming that $z_{i} \in \{1, \dots, K\}$, we may define a transition matrix $P\left(z_{t + 1} = k^\prime \vert z_{t} = k\right)$ with elevated weights along the diagonal, encouraging a “stickiness” in states over time. If we use Gaussian likelihoods $p\left(x_{t} \vert z_{t} = k\right) = \Gsn\left(\mu_{k}, \Sigma_{k}\right)$, we arrive at a version of the Gaussian mixture model where samples tend to remain in the same component for long stretches of time. Relaxing either the Markov or discreteness assumptions leads to further enriched models. For example, a semi-Markov DAG structure supposes that that future states $z_{t}$ may be modulated by the pattern of several recent states, not simply the current $z_{t}$. If hard assignment to one of $K$ clusters is undesirable, we may adopt a mixed-membership approach, enforcing Markov dynamics for $z_{t}$ lying in the simplex.

We can reason about spatial dependence similarly. For example, imagine modeling a spatial transcriptomics dataset where each sample is a high-dimensional vector of gene expression measurements for an individual cell. We expect that, in spite of the hundreds of genes measured, a $K$-dimensional $z_{i}$ could provide a sufficient description of cell-to-cell variation. In this case, we may seek spatial consistency across neighboring $z_{i}$. This can be accomplished by using a Markov graph, inducing dependence along a spatial grid rather than temporal chain. Alternatively, several recent proposals place a Gaussian Process (GP) prior on $z_{i}$, using $K$ output dimensions and two input (spatial) dimensions \citep{townes2021nonnegative, shang2022spatially, velten2022identifying}. The resulting smoothness in latent variables over space can aid interpretability and performance.


\subsection{Inference}
\label{subsec:inferences}

To draw conclusions from a model, we need to infer plausible distributions for unknown parameters and latent variables. This is a separate task from specifying the generative mechanism, and a variety of Monte Carlo and variational methods have been developed to enable inference across different model structures. In fact, the tools for inference have matured to the point that software is now available for generic inference, fitting models with minimal user effort. Nonetheless, it can be instructive to explore the mechanics of inference techniques, especially when considering strategies to enable more efficient inference through model-specific structure. For this reason, this section begins with an overview of probabilistic programming and then gives an overview of the particle filter, an inference method that is widely used for the types of latent temporal modules discussed in Section \ref{subsubsec:structured_latents}.

\subsubsection{Probabilistic programming}

Probabilistic programs are the stochastic analogs of usual, deterministic computer programs. Hence, they are naturally suited to Monte Carlo, and a number of packages have used this fact to support inference in freely-specified generative models. In probabilistic programming packages, both the deterministic and stochastic relationships between parameters and data are represented computationally, and a generic inference algorithm is used to approximate the distribution of all unobserved random nodes (parameters and latent variables), conditionally on the observed ones (data). 

For example, {\tt Stan}, {\tt Pyro}, and {\tt Anglican} provide interfaces for specifying generative mechanisms in this way \citep{carpenter2017stan, bingham2019pyro, wood2014new}. In the background, they parse the conditional dependence relationships to perform various forms of inference, like Markov Chain Monte Carlo (e.g., Hybrid Monte Carlo, the No U-Turn Sampler, the Gibbs sampler), Variational Inference, Importance Sampling, and Sequential Monte Carlo. The result of any of these inference algorithms is a fitted generative model, which can be used to sample from the approximate conditional distribution of unobserved nodes. Critically, to solve a statistical problem using generative models no longer requires manual derivation of model-specific inference algorithms, which historically made the generative approach inaccessible to all but technical specialists.

Though they do not require manual derivations, skillful use of these platforms requires practice. Hands-on introductions can be found at \citep{clark2016bayesian, mcelreath2020statistical, Blau2021, Pyro2022}. Probabilistic programming languages represent distributions as abstract objects, each with their own methods for evaluating distribution-specific properties, like the log probability of a dataset. Though this additional layer of abstraction can seem foreign at first, it allows models to be constructed by manipulating distributions at a high-level. The benefits of learning to operate at this level are analogous to those gained by learning languages that free the programmer from managing low-level memory — more attention can be invested in the design and experimentation rather than the implementation phase.

Here we will give details on an example that illustrates the algorithmic machinery that operates in the background of these inference packages. 
\subsubsection{Particle filter}
\label{subsubsec:pf}

Particle filters are a widely used collection of stochastic models, and since they bridge different ``generative schools," we pause to give a somewhat detailed discussion.
We focus on the latent Markov model of Figure \ref{fig:lr_dag}d. Having observed a sequence $x_{1:T}$, we aim to characterize the unobserved process in states $z_{1:T}$ giving rise to the observed stochastic function as data. 

For example, we may have observed a sequence of economic indices over time. We imagine that the indices $x_{1:T}$ can be succinctly described by latent states $z_{1:T}$, and we are especially curious about when this latent state undergoes rapid transitions. For example, these periods of transition could be interpreted as a period of rapid expansions of specific industries or deteriorations in overall economic equality \citep{kim2017state}.

To this end, we need access to $p\left(z_{t} \vert x_{1:t}\right)$, but in general settings, there is no analytical solution. In this case, it is natural to consider a Monte Carlo approach. If we can sample $z_{t} \vert x_{1:t}$ through some mechanism, then the resulting empirical distribution should approximate $p\left(z_{t} \vert x_{1:t}\right)$ when the number of samples is large enough. These samples are called ``particles,’’ and they play a role analogous to agents in an ABM.

The idea of importance sampling is to use a proposal $\pi \left(z_{1:t} \vert x_{1:t} \right)$ to come up with candidate samples $z_{1:t}^{b}$ from $p\left(z_{1:t} \vert x_{1:t}\right)$. For example, a common proposal is to sample according to the latent dynamics $\Pi\left(z_{1:t} \vert x_{1:t}\right) := \pi_{t^\prime = 1}^{t} p\left(z_{t^\prime} \vert z_{t^\prime - 1}\right)$. Even though it is not possible to sample from the density $p\left(z_{1:t} \vert x_{1:t}\right)$, we assume that it \textit{is} possible to evaluate it for any given $z_{1:t}^{b}$. This makes it possible to upweight candidates with high probability under the true density. Specifically, by using weights $w\left(z_{1:t}^{b}\right) \propto \frac{p\left(z_{1:t} \vert x_{1:t}\right)}{\pi\left(z_{1:t} \vert x_{1:t}\right)}$, the weighted average $\sum_{b = 1}^{B}f\left(z_{1:t}^{b}\right)w\left(z_{1:t}^{b}\right)$ estimates $\Esubarg{p\left(z_{1:t}\vert x_{1:t}\right)}{f\left(z_{1:t}\right)}$.

While it is convenient to sample entire trajectories $z_{1:t}^{b}$ using a single set of weights, this approach deteriorates for large $t$. Only a small pocket of trajectories $z_{1:t}$ in the $t$-dimensional space will have large enough probability, leading to skewed weights. Moreover, since our goal was only to sample $z_{t} \vert x_{1:t}$, there is no need to sample entire trajectories; the marginals at each timepoint are enough. This suggests an alternative to importance sampling, called the particle filter \citep{doucet2001introduction, andrieu2004particle, crisan2002survey}. First, note the recursive relationship between $p\left(z_{1:t} \vert x_{1:t}\right)$ across $t$,
\begin{align*}
p\left(z_{1:t} \vert x_{1:t}\right) &\propto p\left(x_{t} \vert x_{1:\left(t - 1\right)}, z_{1:t}\right) p\left(z_{1:t} \vert x_{1 : \left(t - 1\right)}\right) \\
&= p\left(x_{t} \vert z_{t}\right) p\left(z_{t} \vert z_{t - 1}\right)p\left(z_{1:\left(t - 1\right)} \vert x_{1:\left(t - 1\right)}\right).
\end{align*}
The first term in the final expression is the likelihood of the current observation given the current latent state, the second reflects the transition dynamics, and the third is the previous iteration of the quantity of interest.

If we use a proposal that satisfies $\pi\left(z_{1:t} \vert x_{1:t}\right) = \tilde{\pi}\left(z_{t} \vert z_{1:\left(t - 1\right)}, x_{1:t}\right)\pi\left(z_{1:\left(t - 1\right)} \vert x_{1:\left(t - 1\right)}\right)$, then we also obtain a recursion for the weights,
\begin{align*}
w\left(z_{1:t}^{b}\right) = \frac{p\left(x_{t} \vert z_{t}^{b}\right)p\left(z_{t}^{b} \vert z_{t - 1}^{b}\right)}{\tilde{\pi}\left(z^{b}_{t} \vert z_{1:\left(t - 1\right)}^{b}, x_{1:t}\right)} w\left(z_{1:\left(t - 1\right)}^{b}\right).
\end{align*}

The particle filter uses these intermediate weights to focus in on promising regions of the trajectory space. Specifically, suppose we have a sample of $B$ plausible trajectories $z_{1:\left(t - 1\right)}^{b}$ up to time $t - 1$. Using $\tilde{\pi}\left(z_{t} \vert z_{1:\left(t - 1\right)}, x_{1:t}\right)$, propose one-step extensions $z_{1:t}^{b,\ast}$. Rather than directly using weights $w\left(z_{1:t}^{b,\ast}\right)$ in an importance sampling step, we use these weights to draw,
\begin{align*}
\left(n_{1}, \dots, n_{B}\right) &= \Mult\left(B, \left(w\left(z_{1:t}^{1,\ast}\right), \dots, w\left(z_{1:t}^{B, \ast}\right)\right)\right).
\end{align*}
Trajectories $z_{1:t}^{b,\ast}$ with low weight may have $n_{b} = 0$. In contrast, those with high weight might have large $n_{b}$. To construct an extended set of trajectories $\left(z_{1:t}^{b}\right)_{b = 1}^{B}$, we simply make $n_{b}$ copies of each $z_{1:t}^{\ast, b}$. It turns out that this is exactly a draw from $p\left(z_{t} \vert x_{1:t}\right)$ — no additional reweighting step is necessary, it is already accounted for in the multinomial sampling.

A useful interpretation is to consider the $\left(t - 1\right)^{st}$ coordinate of the partial trajectory $z_{1:\left(t - 1\right)}^{b}$ as one of $B$ particles living in a latent space. Our extension to $z_{1:t}^{b, \ast}$ propagates these particles to the next timestep; the particles also got larger or smaller depending on the weights $w\left(z_{1:t}^{b,\ast}\right)$. In the multinomial sampling step, small particles are eliminated. Large particles are split into several at the same location. The paths that these reproducing particles trace out over time defines a collection of plausible trajectories in the latent space. In this way, the incremental evolution of a collection of discrete particles supports analysis of an otherwise intractable analytical problem. Returning to the economics example, the particle filtering view is to understand the latent dynamics of the economy by simulating many alternative histories, ensuring that at each step, the simulated history is not too inconsistent with the observed data.

\section{Goodness-of-Fit}
\label{sec:calibration}

The difference between successful and failed modeling efforts is rarely the initial analysis. These first attempts are typically disappointing — progress is made by recognizing and addressing limitations. Ideally, this refinement can be guided by formal, automatic processes. Indeed, automation is in some ways responsible for the success of classical goodness-of-fit testing — it made these methods broadly accessible, and even for statistical experts, formal goodness-of-fit evaluation helps in navigating ambiguous settings and enhances reproducibility \citep{kempthorne1967classical}. Classical goodness-of-fit tests have a relatively narrow scope, but many of the core ideas can be extended to the general generative framework. The modular structure of the generative approach makes them naturally amenable to iterative improvement. This section describes progress in generative models that helps bring some of the precision and routine of classical goodness-of-fit testing to more complex, generative settings.

In this context, it is important to develop quantitative and visual techniques for measuring the discrepancy between real with model-simulated data \citep{friedman2001greedy, kale_hypothetical_2019}. Informally, we can compute reference distributions for any number of test statistics and mechanisms. For example, in the single-cell genomics community, packages are available that apply a battery of metrics, like gene-level dispersion and sparsity, to gauge simulator faithfulness \citep{crowell2021built, cao2021benchmark}. 

A key observation is that, rather than specifying the discrepancy measure in advance, a discriminator can itself be learned from the data. This is the approach of \citep{friedman2003}, who uses ensemble models to distinguish real and simulated data. To calibrate the typical effectiveness of the discriminator, it can be useful to contrast pairs of sample from the same generative mechanism; this serves as the analog of the reference distribution in classical testing. An ineffective discriminator is used as evidence of goodness-of-fit — if the real data cannot be distinguished from samples from a given generative mechanism, that mechanism must approximate reality well. Moreover, if a discriminator is able to distinguish between real and simulated data, then the regions for which it is most accurate can be used to guide refinements to the generative mechanism.

\subsection{Example: Mixture modeling}
\label{subsec:mixture_modeling}

\begin{figure}
    \centering
    \includegraphics[width=.5\textwidth]{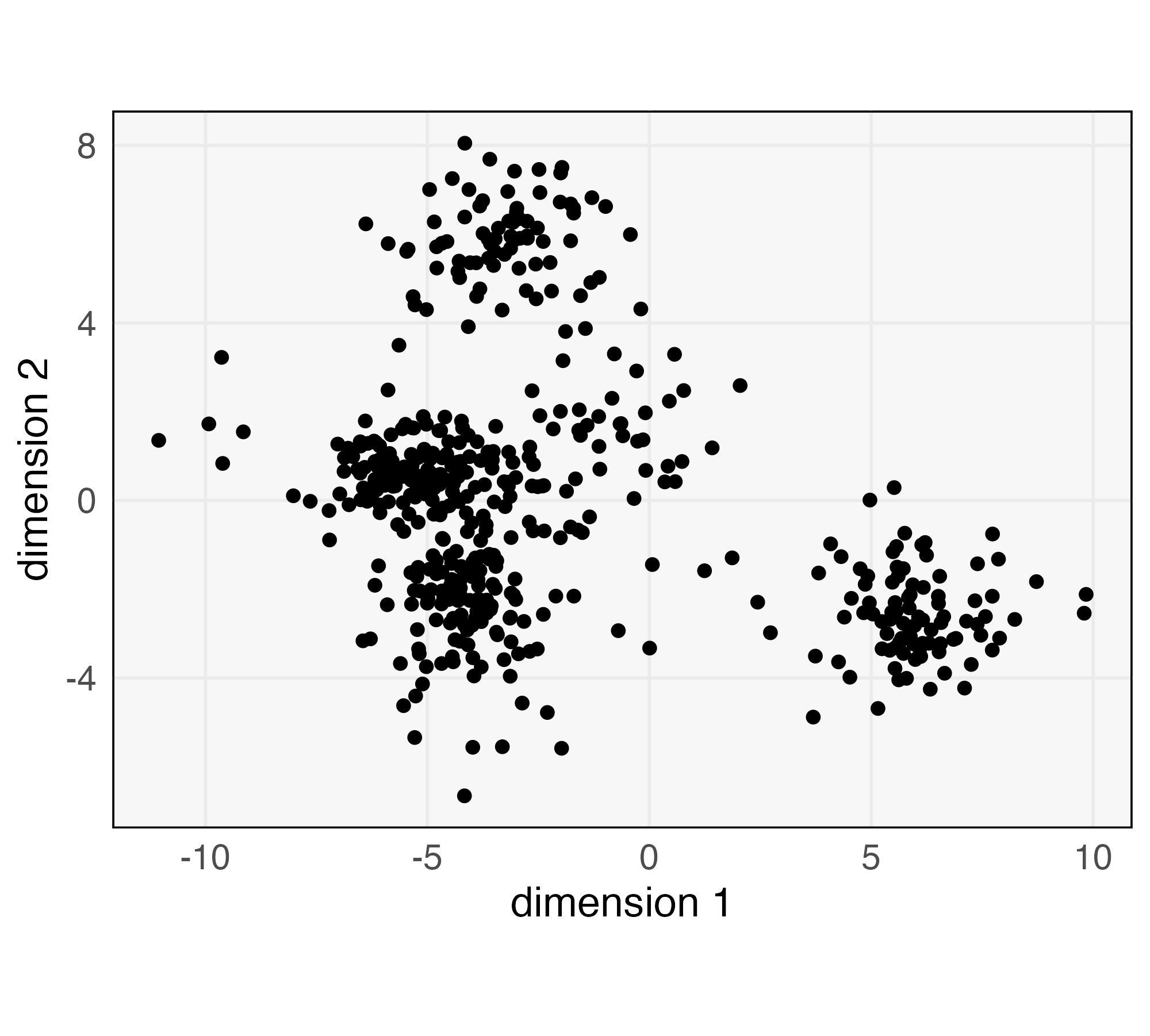}
    \caption{The mixture dataset of interest in Section \ref{subsec:mixture_modeling}.}
    \label{fig:true_mixture}
\end{figure}

We next provide an example of iterative model building using a discriminator. Though the example dataset is small, it contains enough complexity to underscore the value of quantitative goodness-of-fit evaluation. Imagine we have been presented with the data from Figure \ref{fig:true_mixture}.  Noting the clear mixture structure and fairly elliptical shape in each component, we decide to fit a Gaussian Mixture Model with $K = 4$ components. We place a Gaussian prior on the means and assume a diagonal covariance with equal variance across both dimensions and shared for all components.

\begin{figure}
    \centering
    \includegraphics[width=\textwidth]{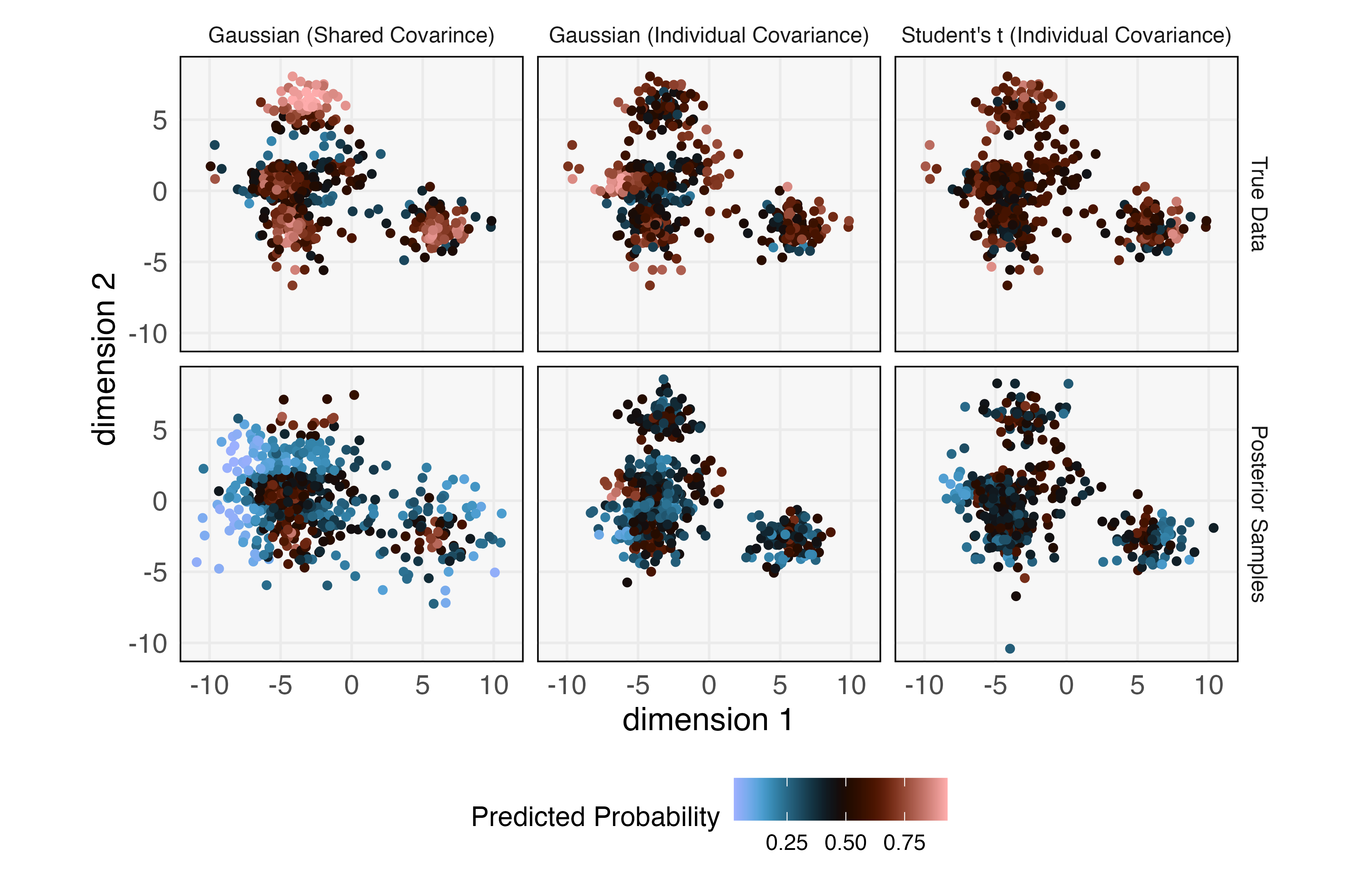}
    \caption{Discrimination probability across a sequence of models, arranged from left to right. The true dataset is given in the top row, simulated posterior predictive samples from each model appears in the bottom row. Points are colored in by their predicted probability of belonging to the true or simulated dataset, derived from a GBM discriminator. As the simulated data appear to more effectively capture structure in the observed data, the discriminator struggles to identify any difference. }
    \label{fig:modeling_iteration}
\end{figure}

To evaluate goodness-of-fit, we fit a Gradient Boosting Machine (GBM) discriminator to distinguish between true and simulated data \citep{friedman2001greedy, friedman2003}. The GBM is a flexible, tree-based method that can learn nonlinear boundaries between classes. Evaluating classifier accuracy with the same data used during training would not allow us to detect overfitting, so we use 10-fold repeated cross validation. This randomly assigns each of the real and simulated samples to one of ten folds. Then, ten models are trained, each one holding out one fold, which is used to evaluate accuracy. The full process is repeated ten times, and all the fold-level holdout accuracies are averaged. Through this process, we find that the model achieves an average 65.5\% holdout accuracy, indicating substantial room for improvement. To understand the source of this ability to discriminate between true data and simulated samples, the left panel of Figure \ref{fig:modeling_iteration} visualizes each simulated and real sample with its predicted class probability overlaid. An over-represented region near the top left is visible -- this is the bright blue region in the large cluster in the bottom left panel. Since these points are blue, the GBM has learned that samples in this region are almost always simulated, not real. It seems that two clusters have been merged. Further, we observe that the density of points in the center of the bottom right cluster is higher in the real relative to the simulated data. Indeed, the predicted probability of the true data class within the core of this component is noticeably higher than the corresponding probability for simulated data.

To address these issues, we implement two changes to the model,

\begin{itemize}
\item Since our model seems to be incorrectly merging clusters, we increase $K$ to 5. 
\item One explanation of the over-represented core in the bottom right cluster is that the real data have smaller variance in that cluster. Hence, we allow each mixture component and dimension to have its own variance, though the covariance is still diagonal.
\end{itemize}

We again fit a GBM to distinguish simulated and real data. For this model, the average holdout accuracy is now 55.5\%. The corresponding predicted probabilities are shown in the middle panel of Figure \ref{fig:modeling_iteration}. The performance of the discriminator has deteriorated, indicating an improved model fit. Further, the failure to cover all true data has been alleviated. However, the core of one of the mixture components continues to be over-represented, now in the bottom left cluster.

We next realize that an alternative explanation for the underrepresented cores may be the presence of outliers. Indeed, an outlier has the potential to inflate the estimated variance within a cluster, leading to a larger spread for the simulated data. Therefore, we consider fitting a mixture of $t$-distributions instead. As in the previous iteration, we use $K = 5$ clusters and diagonal (but not necessarily equal) covariances. A GBM discriminator applied to the results of this model now only achieves a holdout accuracy of 50.5\%. Discriminator predicted probabilities in the far right panel of Figure \ref{fig:modeling_iteration} suggest that the problem of over-representing the core of certain mixtures has disappeared. The only potential issue is the presence of some regions close to outliers where the simulator generates unnecessary samples. Overall, the fit is noticeably improved relative to either of the two previous attempts, and this is reflected in the discriminator’s essentially random guesses.

\subsection{Discrepancy measures}

By this point, it should be no surprise that the true data are a mixture of 5 Student-$t$ distributions with unequal, but diagonal covariance matrices. Nonetheless, this toy setup captures many features of real data analysis \citep{blei2010probabilistic, gabry2019visualization}. Rather than proposing an overly complicated initial fit, it is natural to begin with a textbook model, but there are bound to be mistakes along the way (e.g., mis-specifying the number of clusters above). Since generative models support simulation, it is possible to train a discriminator to hone in regions of inadequate model fit. Here, we followed the discriminator training of \citep{friedman2003}, applying a nonlinear, tree-based model to transform a goodness-of-fit problem into a matter of classification. Specifically, the accuracy of a classifier provides a measure of discrepancy between true and simulated samples. Low classification accuracy suggests that the generative model is capable of mimicking even subtle patterns in the real data. The advantage of this overall perspective is that now any prediction algorithm can be turned into a measure of distributional discrepancy. Based on the guidance from a discrepancy measure, the modeler can introduce necessary complexity, adapting dependence structures and distributions until a satisfactory fit is found.

Goodness-of-fit evaluation is traditionally viewed as a hypothesis testing problem, but the example above highlights the close relationship between testing and classification. Indeed, this connection has sparked renewed interest in evaluation for more complex generative modeling settings. We note in particular the Stein Discrepancy approach \citep{gretton_kernel_2012, gorham2019measuring}. Given a target $\P$, whose distribution we imagine the data may belong to, it is often possible to construct a “Stein Operator” $\mathcal{T}$ and test function class $G$ such that, whenever $x_{i} \sim \P$, the sample averages $\frac{1}{n}\sum_{i=1}^{n} \mathcal{T}g\left(x_{i}\right) \approx 0$ for any $g \in \mathcal{G}$. The collection of functions $\{\mathcal{T}g : g \in G\}$  can be viewed as a collection of test functions, and whenever the observed values of these test functions is far from 0, we have an indication that the true distribution of $x_{i}$ is not $\P$. Formally, the Stein Discrepancy is defined as the maximum expected value over all test functions, $\sup_{g \in G} \|\Earg{\mathcal{T}g\left(X\right)}\|$.

The value of this approach is that operators $\mathcal{T}$ and classes $G$ can be constructed even for distributions whose normalizing constants are intractable. This makes them especially attractive for evaluation in the context of generative models. They make it possible to leverage the known unnormalized component of the density and avoid generic goodness-of-fit tests, which often have low power, since they must be valid across wider classes of densities.

\section{Simulation and emulation}
\label{sec:emulation}

Simulations are a general problem-solving device: A good simulator encodes beliefs about a system and enables decision-makers to query a wide variety of states and perturbations, even those that were never observed. They are especially useful in complex systems where dynamic spatio-temporal data make analytical calculations intractable. 
Decision-making can follow the system from the individual agents up to the emergence of macro-properties which then can be evaluated using sensitivity and stability indices. Conclusions can be drawn by observing the evolution of units across repeated trajectories within and across parameter settings.

For example, a variety of simulators were used to compare the potential effects of alternative non-pharmaceutical interventions in the COVID-19 pandemic \citep{kerr2021covasim, ferguson2020impact, hinch2021openabm}. Guided by expert knowledge, these simulations allowed a  critical evaluation of alternative paths through the pandemic. For example, university presidents could ask the simulator how switching to remote classes for variable numbers of weeks would influence the total number of cases over the semester or how changing the number of students in each dorm might influence the total number of COVID-19 clusters.

These simulations can be guided by local, agent-level rules, as in ABMs. When modeling the spread of a contagion, the modeler decides on the structure of agent-level interactions and the probability of disease spread given an interaction. Since these mechanisms are defined by programs, not explicit mathematical densities, it is not possible to explicitly evaluate the probability of a simulated dataset. In the absence of formulas, we might be resigned to informal deductions based on tinkering with parameters in the simulation mechanism. 

However, a flurry of recent activity, surveyed below, demonstrates that this informal decision-making can be formalized into statistically-guaranteed inference. Simulation and inference are closely linked — whenever it is possible to simulate data, it is possible to use an observed dataset to infer plausible simulation parameters. This is true even when the simulation mechanism is described by local, agent-level rules. It is possible to map local rules to global properties of the simulated systems, and if variations in local rules are reflected in global properties, then reversing the mapping supports inference. These methods support statistical inference in situations, like contagion modeling, where the mathematical form of the overall probabilistic mechanism may be difficult to express, but where it is natural to define local computational rules that govern the system.

\subsection{Approaches}

A simulation is governed by global parameters $\theta$, is initialized in a state $x_0$, and evolves across iterations $t$. The state of the system refers to all of its variables at a given time. The current state $x_t$, the historical trajectory of the system $\left(x_{s}\right)_{s = 0}^{t - 1}$, and random noise can all govern the evolution of $x_t$ into $x_{t + 1}$. We assume the simulation is stopped after $T$ steps, and write the full trajectory of states as $\*x := x_{1:T}$.

Many problems reduce to answering the following questions,
\begin{itemize}
\item Hypothetical outcomes: For a given parameter $\theta$, summarize the distribution of some statistic $S\left(\*x\right)$  across simulations evolving according to $\theta$. This can be approximated by running an ensemble of simulations $\*x^{b}$ for $b = 1, \dots, B$ and computing the summary $S$ across the ensemble.
\item Parameter inference: Given a series of observed states $\*x$, determine a distribution of plausible parameters $\theta$ of the simulator.
\end{itemize}

The methods below have been developed to address these questions. 

\subsubsection{Approximate Bayesian Computation}
\label{subsubsec:abc}

Approximate Bayesian Computation (ABC) was one of the earliest approaches to likelihood-free inference, and it remains widely used in practice \citep{tavare1997inferring, pritchard1999population, sisson2018handbook, beaumont2019approximate}. It is used to answer parameter inference questions. Let $F\left(\*x \vert \theta\right)$ denote the (inaccessible) probability density over trajectories $\*x$ for simulators parameterized by $\theta$, and define a proposal $\pi\left(\theta\right)$ over the simulator’s parameter space $\Theta$. Generate an ensemble of simulated datasets, where parameters are themselves drawn from the proposal,
\begin{align*}
\*x^{b} \vert \theta_{b}&\sim F\left(\*x \vert \theta_{b}\right), b = 1, \dots, B \\
\theta_{b} &\sim \pi\left(\theta\right)
\end{align*}
To compare a pair of datasets, compute a distance between sufficient statistics, $d\left(S\left(\*x\right), S\left(\*x^\prime\right)\right)$. Datasets with the same sufficient statistics are considered equivalent. Given a real dataset $\*x$, ABC induces a posterior on $\theta$ by building a histogram from all those $\theta_{b}$ where $d\left(S\left(\*x^b\right), S\left(\*x\right)\right) < \epsilon$. The parameter $\epsilon$ governs the quality of the approximation — small $\epsilon$ leads to more faithful approximations, but results in most forward simulations being omitted from the posterior histogram’s construction.

Though simple to implement and applicable across a variety of settings, ABC can be greedy, requiring many forward simulations $B$ before arriving at useful posterior approximations. Moreover, estimates are sensitive to the choice of $S$ and $\epsilon$ \citep{prangle2018summary}. For this reason, we turn next to alternatives that make stronger assumptions on the relationship between $\Theta$ and $x$, but which are more sample efficient.

\subsubsection{Gaussian Process Surrogates}

For computationally intensive simulations, it can be prohibitive to generate more than just a few runs. To explore a wide range of configurations $\theta$ in an efficient way, it is often possible to estimate a surrogate model $\tilde{y}\left(\theta\right)$ of the relationship $y\left(\theta\right)$ between $\theta$ and summaries $S\left(\*x\right)$ of interest. This new surrogate supports more rapid exploration of hypothetical outcomes, compared to running the entire simulation for each choice $\theta$. The main assumption is that $y$ is smooth; i.e., simulations using similar values of global parameters $\theta$ are expected to have similar summary statistics.
Surrogate models are often estimated using either normalizing flow \citep{brehmer2020mining} or Gaussian Process (GP) \citep{gramacy2020surrogates, baker2022analyzing} models. Both approaches have the advantage that they 
emulate the likelihood $F\left(\*x \vert \theta\right)$, supporting more direct inference over $\theta$ than ABC \citep{cranmer_frontier_2020, dalmasso_likelihood-free_2021}.

We detail the GP approach to simulation surrogates. This approach has been successfully applied to Ebola contagion modeling, fish capture-recapture population estimation, and ocean circulation models -- see~\citep{baker2022analyzing} for a comprehensive review. Suppose we have computed $B$ simulation runs $\*x^{b}$ for parameters $\theta_{b}$. To ease notation, write $y_{b} = S\left(\*x^{b}\right)$. For the moment, suppose that $y_{b} \in \reals$ is one-dimensional and that we aim to approximate $y^{\ast} := S\left(\*x^{\ast}\right)$ at a new configuration $\theta^\ast$. A GP surrogate assumes a prior $\left(y_{1}, \dots, y_{b}, y^\ast\right) \sim \Gsn\left(0, C\left(\theta\right)\right)$, where $C\left(\theta\right)$ is a covariance matrix depending on the simulation parameters via a pre-specified covariance function. For example, the $\left[C\left(\theta\right)\right]_{bb^\prime}$ could be $C\left(\theta_{b}, \theta_{b^\prime}\right) = \gamma\exp{-\kappa\|\theta_{b} - \theta_{b^\prime}\|^2}$ for hyperparameters $\gamma, \kappa$. To emulate the simulator at a new parameter $\theta^\ast$, we compute the posterior $y^{\ast} \vert \left(y_{b}\right)_{b = 1}^{B}$, which is available in closed form by the rules of Gaussian conditioning. In particular, the mean and variance of $S\left(\*x^\ast\right)$ can be immediately derived.

In the case of multidimensional $y_{b}$, several options are available. First, we may model each coordinate of the summary with a separate GP. Alternatively, we may swap dimensions between the output and input space. For concreteness, suppose that $\left(y_{1}^b, \dots, y_{T}^{b}\right)$ is a $T$-dimensional summary of the temporal evolution of the simulation at configuration $\theta_{b}$. We can define a covariance function $C\left(\left(\theta, t\right), \left(\theta^\prime, t^\prime\right)\right)$ that correlates
simulation outputs with similar parameters $\theta$ and timepoints $t$. This new covariance can be used in a GP model as before, and the conditional 
probability $y_{1}^{\ast}, \dots, y_{T}^{\ast} \vert \left(y_{t}^{b}\right)_{t, b = 1}^{T, B}$ 
can be found for any new sequence of outputs $\left( y_{1}^\ast, \dots, y_{T}^\ast\right)$.

\subsection{Examples}

\subsubsection{Batesian mimcry}

We next illustrate the use of simulation-based inference in a simple ABM from evolutionary genetics. The ABM provides an environment for reasoning about Batesian mimicry, a phenomenon where species evolve similar appearances under selection pressures. A classical example is the resemblance between viceroy and monarch butterflies. Birds are known to avoid monarch butterflies, whose diet includes plants toxic to birds. Viceroys are edible, but when they happen to have similar appearance as monarchs (black and red wings), birds avoid them.

This story can be captured by an ABM model, one that is implemented in the NetLogo package's built-in library \citep{tisue2004netlogo}. This ABM has three types of agents — monarchs, viceroys, and birds. Monarch and viceroy butterflies have a categorical “color” attribute with 20 possible levels. This represents how they appear to birds. Birds have vector-valued “memory” attribute, storing colors of up to three of the most recently eaten monarchs. All agents have an $\left(x, y\right)$ location attribute. At each time step, nearby birds and butterflies interact. If the butterfly’s color is contained within the bird’s memory, the butterfly survives; otherwise it is eaten. If the eaten butterfly happened to be a monarch, then the monarch’s color is added to the bird’s “memory” vector. If the vector is now longer than three, then the oldest color is removed from memory. Also at each time step, butterflies replicate with a small probability. When they replicate, they either create an exact replicate or they create a “mutant” with a color attribute drawn uniformly from one of the 20 options. The probability of creating a mutant is controlled by a global mutation rate.

The salient feature of this model is that the average colors for the two species eventually converge to one another.  In this way, local rules are able to capture the emergent phenomenon of Batesian mimcry. The fact that viceroys survive longer when they have colors that overlap with monarchs creates a selection pressure that results in the species appearing indistinguishable, on average.

\begin{figure}
    \centering
    \includegraphics[width=0.9\textwidth]{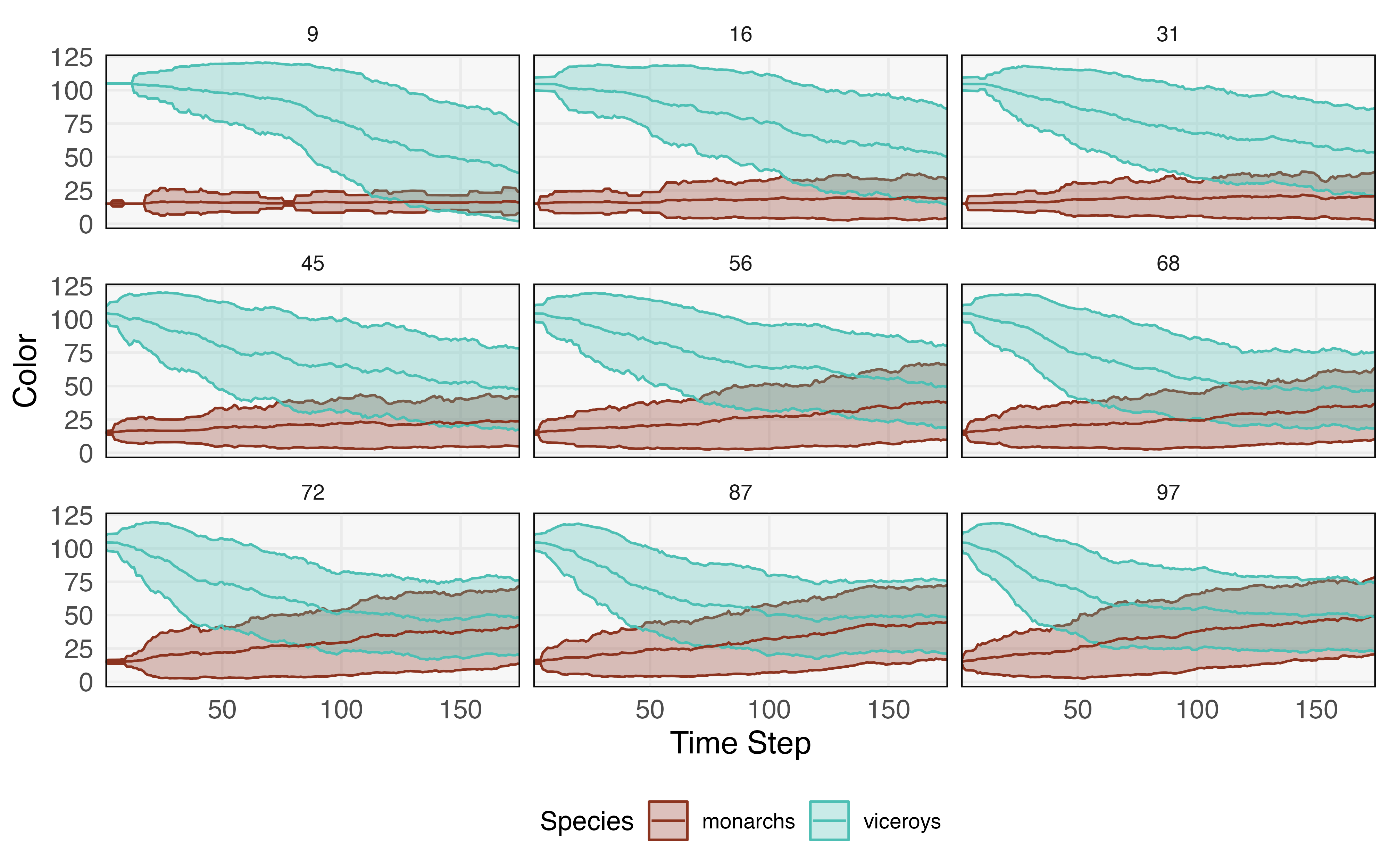}
    \caption{Examples trajectories of butterfly color in the Batesian mimcry ABM. Colors encode butterfly species. Solid lines and envelopes give the average color across all butterflies from that species over time, plus-or-minus one standard deviation. Panel titles give global mutation rate, and are sorted by slowest to fastest evolution.}
    \label{fig:mimicry_trajectory}
\end{figure}

We next pose an inference question: Given an observed simulation trace, can we infer the mutation rate? Figure \ref{fig:mimicry_trajectory} shows that higher mutation rates lead to more rapid convergence, providing a basis for inference. We approach the problem using a variant of ABC based on the particle filter called Sequential Monte Carlo  - Approximate Bayesian Computation (SMC-ABC) \citep{del2012adaptive, jabot2013easy}, a blend of the algorithms discussed in sections \ref{subsubsec:pf} and \ref{subsubsec:abc}. This requires specification of a prior on mutation rates and a summary statistic to use for the basis of similarity comparisons. We choose a uniform prior over mutation rates from 0 to 1. Our summary statistic is defined as $S\left(\*x\right) = \left(\left(\bar{x}^{\text{viceroy}} - \bar{x}^{\text{monarch}}\right)_{1:\frac{T}{4}}, \dots, \left(\bar{x}^{\text{viceroy}} - \bar{x}^{\text{monarch}}\right)_{\frac{3}{4}T:T}\right)$ giving the difference in average color for viceroy and monarch butterflies over four equally spaced time intervals.

\begin{figure}
    \centering
    \includegraphics[width=0.9\textwidth]{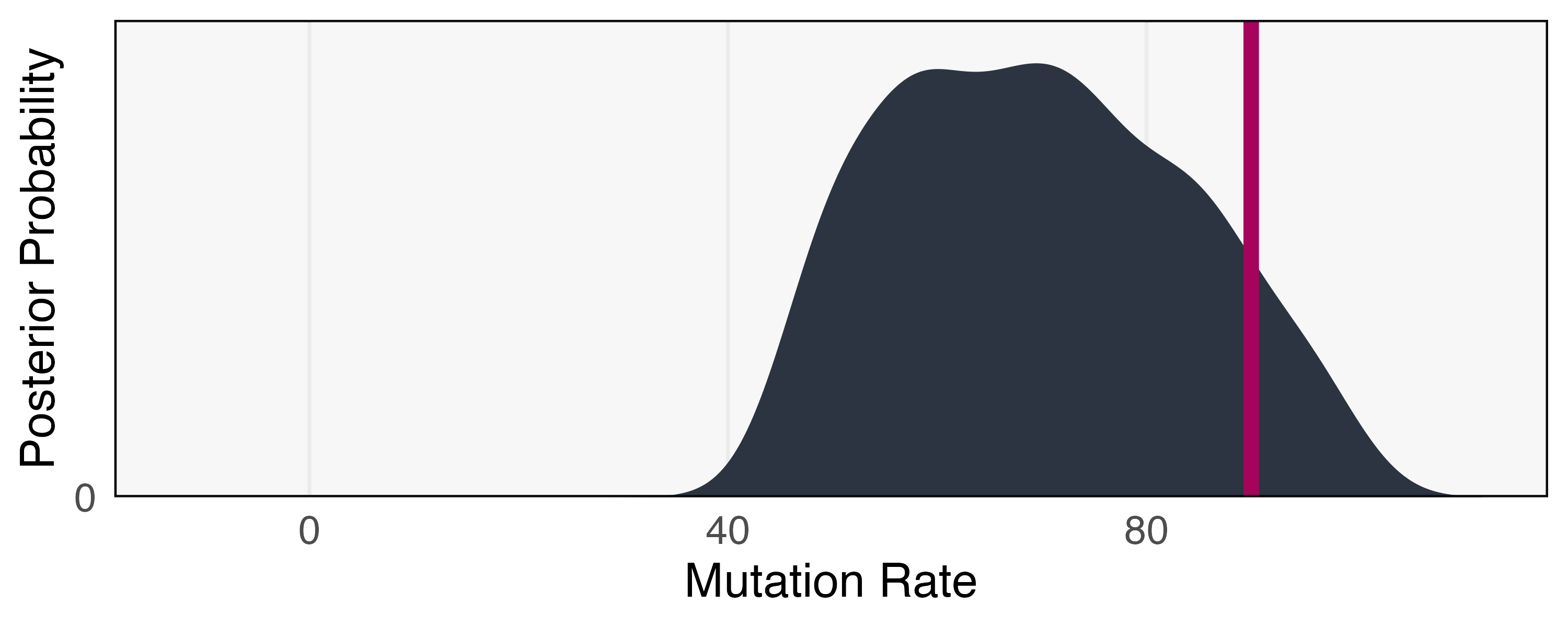}
    \caption{The posterior probability of mutation rates in a run of the Batesian mimcry ABM when using a true rate of 90. By using Sequential Monte Carlo with Approximate Bayes(SMC-ABC), posterior probabilities can be derived without access to a true likelihood. Particles and weights from SMC have been smoothed to form a posterior density.}
    \label{fig:mutation_inference}
\end{figure}

Our ``real'' data summary statistic is found by simulating one run with a mutation rate of 90. We run each simulation for $T = 175$ time steps and use a summary statistic tolerance of $\epsilon = 5$. We continue simulating until 250 samples are accepted. The SMC-ABC posterior is shown in Figure \ref{fig:mutation_inference}. Our inference has clearly ruled out any mutation rates below 40, and though the posterior mode is near the true value, there remains high uncertainty. This is consistent with the difficulty in differentiating large mutation rates visually in Figure \ref{fig:mimicry_trajectory}. We note that this posterior is still approximate, and lowering the tolerance and increasing the number of required posterior samples would lead to a more precise estimate, albeit at the cost of increased computation.

\subsubsection{COVID-19 nonpharmaceutical interventions}

Our next example describes how Gaussian process surrogates can support efficient decision-making when working with a compute intensive COVID-19 simulator. The simulator we use, Covasim, is an ABM originally developed to support reasoning about hypothetical outcomes of various nonpharmaceutical interventions to COVID-19, including workplace and school closures, social distancing, testing, and contact tracing \citep{kerr2021covasim}. Each agent represents an individual, and their state reflects the potential progression of disease, from asymptomatic-infectious to various degrees of disease severity, and finally to recovery or death. At each time step, agents interact across a predefined set of networks, representing family, school, workplace, or random contact networks, and during these interactions, the disease can spread.

Alternative interventions inhibit these interactions in different ways -- we focus here on testing and contact tracing interventions. These are parameterized by $p_{\text{test}}$, $n_{\text{test}}$, and $p_{\text{trace}}$. $p_{\text{test}}$ modulates the fraction of symptomatic individuals who receive a COVID-19 test. If their test comes back positive, the agents enter a self-imposed quarantine, during which they do not transmit the disease. The intensiveness of the contact tracing effort is controlled by two parameters, the number of days $n_{\text{test}}$ needed before contact tracing notifications are made, and the probability $p_{trace}$ that a contact of positively-tested agent is correctly traced.

\begin{figure}
    \centering
    \includegraphics[width=0.95\textwidth]{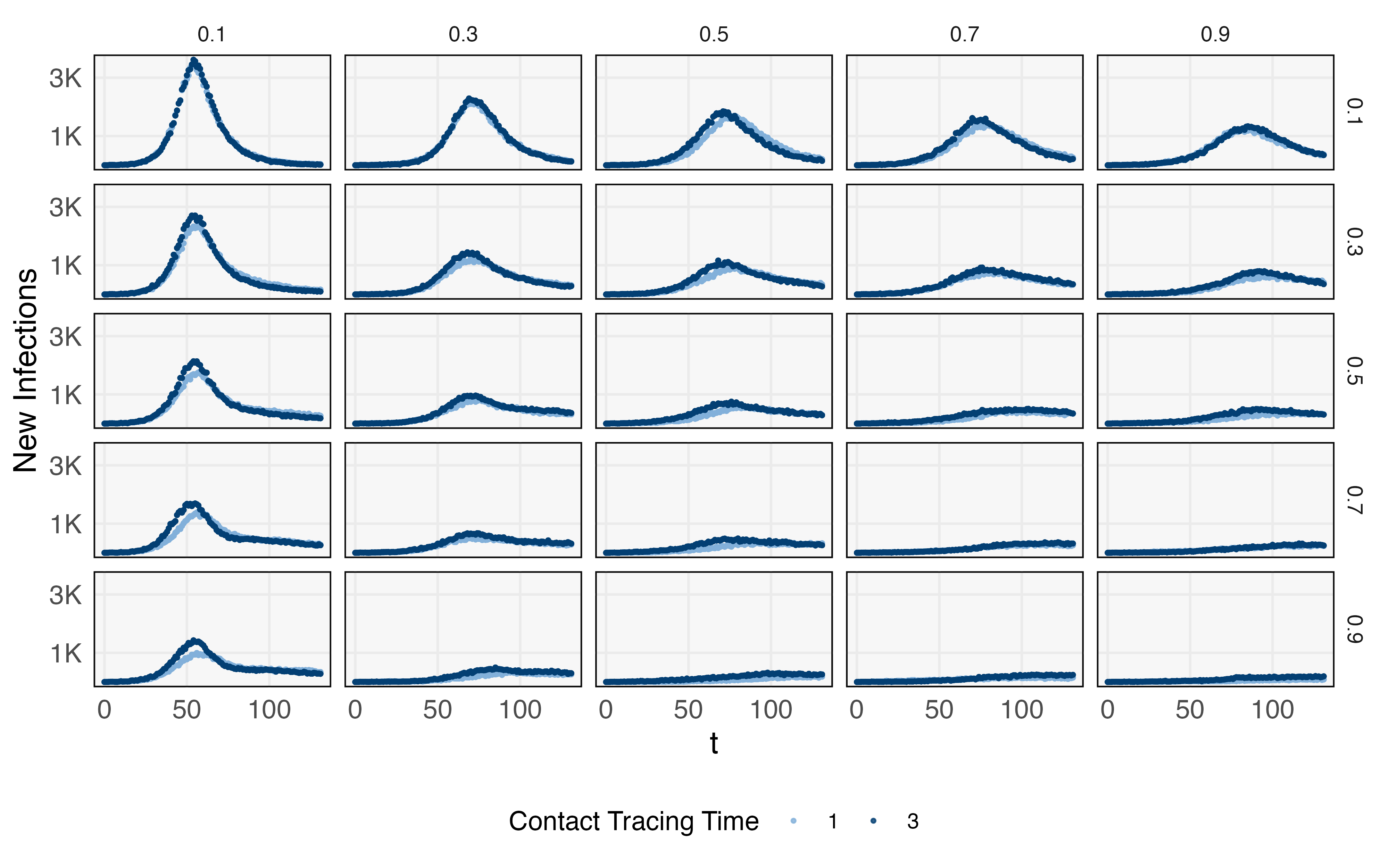}
    \caption{Hypothetical outbreak trajectories from the Covasim ABM when varying three simulation parameters. Column labels give the probability that a symptomatic agent is given a test. A positive result triggers a quarantine for that agent and contact tracing for its recent contacts. Rows correspond to the probability that a contact of a positively tested agent is traced down and isolates. The number of days required for contact tracing to trigger isolation is encoded by color.}
    \label{fig:covid_trajectories}
\end{figure}

\begin{figure}
    \centering
    \includegraphics[width=\textwidth]{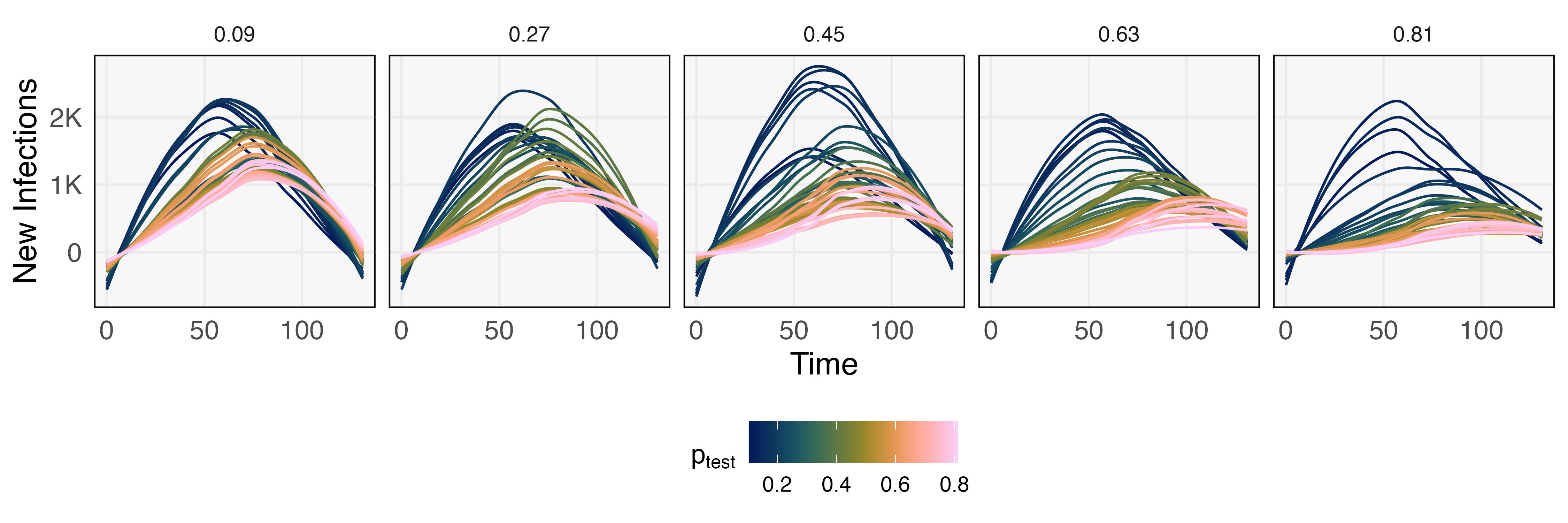}
    \caption{Hypothetical outbreak trajectories under differential testing and contact tracing capabilities,  derived using a Gaussian Process emulator of the Covasim ABM (see Figure \ref{fig:covid_trajectories}). Panels give the probability that an agent that interacted with an agent in an infectious state is both traced down and placed into quarantine. Colors encode the probability that a symptomatic agent is given a test. The qualitative influence of these parameters mirror those in the full simulation, but are derived with less computation.}
    \label{fig:emulator_trajectories}
\end{figure}

Before evaluating the effects of alternative interventions, we first calibrate the simulator to a dataset of daily infections, tests, and deaths, representing the initial phase of an outbreak on a population of size 100,000. This calibration step runs trials with varying link infection and death probability, choosing a combination that best fits the observed initial outbreak. Given these background parameters, we use the simulator to evaluate hypothetical effects of different $p_{\text{test}}$ and $p_{\text{trace}}$ — these effects can be used to set targets for testing and contact tracing. Since evaluating the simulator along a fine grid of these parameters is costly, we train a GP surrogate to a coarser set of simulations, shown in Figure \ref{fig:covid_trajectories}. This initial set of runs already gives the general outlines of intervention effects, and a surrogate model makes it possible to represent the analogous trajectory for any $\left(p_{\text{test}}, p_{\text{trace}}\right)$ combination of interest. For example, Figure \ref{fig:emulator_trajectories} provides 250 trajectories across a range of $p_{\text{test}}$ and $p_{\text{trace}}$ — this takes 2.65 minutes on a laptop with a 3.1 GHz Intel Core i5 processor and 8GB memory. In contrast, computing just one trajectory from the Covasim ABM takes 6.20 seconds, and the estimated time for all 250 is 25.8 minutes.

\section{Conclusion}


Generative models have emerged as key instruments for inferential and hypothetical reasoning. As models, they distill data into meaningful theories, and as simulators, they support inquiry into alternative courses of action. Throughout this review, we have seen that the data they produce can guide the myriad decisions needed for effective design and modeling, from establishing the experimental setup of a genomics study to weighing potential responses to a pandemic. Calibration against observed data grounds reasoning in a concrete setting, while direct control of parameters supports imagination of alternatives.

The practical utility of these approaches has led to their widespread adoption across many communities. This review has drawn examples from physics, economics, epidemiology, behavioral ecology, biostatistics and computer science. Even disciplines far removed from real-world sensors have found value in simulated data; for example, \cite{mazur2008finding} uses generative models to develop and refine mathematical conjectures in number theory. We have noted the difficulties in leveraging methods developed across several fields, but these are often due more to cultural habits and language rather than any fundamental differences in problem solving strategy. Indeed, many of the methods reviewed here are drawn from the broader program of strengthening the ties between communities — discovering ways in which computational simulators can support statistical experimental design, and drawing from concepts in statistical inference to support analysis of ABMs. Generative models are rapidly becoming the lingua franca of science.

Some of the most interesting advances in generative models do more than bridge communities — they enrich our language for reasoning about chance. In particular, we have seen a continuum emerging between granular ``agent-particle'' models and those based on smooth probabilistic densities. In complex simulations, smooth emulations provide a path to tractability. Discrete ABMs can represent complex interactions and heterogeneity, but a smooth surrogate can simplify direct manipulation and queries. In the other direction, when closed-form recursions for posterior distributions in latent Markov models proved impossible to obtain, a shift towards the ``particle'' perspective proved critical for effective inference. There is a spectrum between smooth and discrete representations for stochastic systems, and transitioning across it can be a powerful problem solving device.

It is increasingly the case that generative models are used to support introspection and discovery. Simulation mechanisms make it possible to compare and analyze scientific workflows and data analysis strategies. Recent statistical work has provided strong evidence for specific high-dimensional phenomena based on simulation alone \citep{bertsimas2016best, bertsimas2020sparse, chen2020look, hastie2020best}. Within statistics, distinctions between theoretical and empirical analysis have long been blurred, with both theorems and simulations used to clarify key issues across the discipline. A generative perspective helps these complementary approaches cohere — experiments need conceptual mechanisms, and theory needs observed structure.

Research can hold too many surprises for broad-brush forecasts to be taken too seriously. However, it is often worthwhile to pay careful attention to recent developments, with an eye towards how scientific practice has evolved. In this spirit, consider the statistical activity sparked by our societal need to respond to the COVID-19 pandemic. During the early phases of the pandemic, and indeed with each unexpected development, actions could not be informed by data alone, and decision-making had to be guided by simulation before data could complement the process. Against this backdrop, rapid progress was made in bridging inference with simulation, facilitating both critical evaluation of uncertainties and imagination of hypothetical outcomes. Researchers found that successful problem solving during a crisis requires communication across disciplinary boundaries, with reproducible, transparent models playing a central role. Responses to future crises, in particular those driven by climate and ecological change, will likely require an even greater degree of coordination and model-guided decision-making, and we expect them to be similarly dependent on progress in generative modeling.

The broadened scope of problems that have been considered by generative models poses a challenge on several levels. On a non-technical level, even when two communities use generative models, it is easy for ideas to be lost in translation. For example, what in statistics might be called nonidentifiability has come to be called “sloppiness” by those using simulations in ecology and biology. Similarly, nonlinearity in statistics usually refers to the form of a regression function, while in simulation, nonlinearity is used in the sense of the system being governed by a nonlinear partial differential equation. Care needs to be taken to work on teams with diverse intellectual backgrounds, and to hear out each of the technical ideas without falling into the traps of language or arguments over the ownership of those ideas.

On a technical level, there remains a challenge in structuring the process of model refinement. Models give a form of approximation, and two questions are at the back of every modeller’s mind: (i) is the approximation sufficient and (ii) if not, how can it be improved? In ABMs for example, there are always additional attributes that could be given to each agent. These would make the simulation more realistic, breaking the partial exchangeability implicitly encoded when using only a more limited attribute set. Similarly, for variational methods, we often deliberately ignore conditional dependence relationships for the sake of computational tractability. Computational constraints and our uncertainties about a system often force us to approximate, sometimes dramatically so, and the question becomes: does it matter? 

When working with generative models, we can gauge the approximation through simulation. The data themselves are imbued with enough variability that, if our model generates data that blend in perfectly, then the approximation is “good enough.” This approach hinges critically on access to effective discrepancy measures. Moreover, when discrepancies are found, they should be transparent enough to guide model improvement. We see potential for recent theoretical work with Stein's Method \citep{gorham2019measuring} and optimal transport \citep{solomon2015convolutional} to translate into tools for model criticism and refinement, but at the moment, the practice of generative model building remains more of an art. The flexibility and transparency of generative models that makes them so relevant to modern, collaborative science is also a source of worthwhile challenges for the statistical community. 



\bibliographystyle{ar-style1}
\bibliography{GenerativeReview}
\nocite{*}

\end{document}